\documentclass[11pt]{iopart}
\usepackage{hhline}
\usepackage{iopams}
\usepackage{amsfonts}
\usepackage{epsfig}
\usepackage{bbm}
\usepackage{mathrsfs}
\usepackage{latexsym}
\usepackage{showlabels}
\usepackage{color}
\usepackage{tensor}

\usepackage{accents}
\usepackage{nicefrac}
\usepackage{psfrag}
\usepackage{graphicx}
\usepackage{subcaption}
\usepackage[british]{babel}
\usepackage{afterpage}
\usepackage{fancyhdr}
\usepackage{tikz}
\usetikzlibrary{decorations.pathmorphing}
\usepackage{hyperref} 
\usepackage{mcite} 

\def\bZ{\mathbb{Z}}
\def\bN{\mathbb{N}}

\def\dd{\mathrm{d}}
\def\im{{\rm i}}
\def\H{\mathcal{H}}
\def\O{\mathcal{O}}
\def\C{\mathcal{C}}
\def\implies{\Longrightarrow}

\newcommand{\J}[2]{J_{#1}\left( \frac{\sqrt{#2}}{\hbar}v\right)}
\newcommand{\K}[2]{K_{#1}\left( \frac{\sqrt{#2}}{\hbar}v\right)}

\newcommand{\dl}[1]{\frac{\dd #1}{2\pi\hbar}}
\newcommand{\dk}[1]{\frac{\dd #1}{2\pi}} 
\newcommand{\sgn}[1]{\mathrm{sgn} \left( #1 \right)}
\newcommand{\expval}[1]{\left< #1 \right>}
\newcommand{\eqref}[1]{(\ref{#1})}

\newcommand{\pdv}[2]{\frac{\partial #1}{\partial #2}}
\newcommand{\braket}[2]{\left< #1 | #2 \right>}
\newcommand{\abs}[1]{\left| #1\right|}

\begin{document}
\title{Unitarity, clock dependence and quantum recollapse in quantum cosmology}
\author{Steffen Gielen${^1}$, Luc\'ia Men\'endez-Pidal${^2}$}
\address{%
  ${^1}$School of Mathematics and Statistics,
  University of Sheffield,
  \\
  Hicks Building,
  Hounsfield Road,
  Sheffield S3 7RH,
  United Kingdom
  \\[.5em]
  ${^2}$School of Mathematical Sciences,
  University of Nottingham,
  \\
  University Park,
  Nottingham NG7 2RD,
  United Kingdom
}
\ead{s.c.gielen@sheffield.ac.uk, lucia.menendez-pidal@nottingham.ac.uk}

\begin{abstract}
We continue our analysis of a quantum cosmology model describing a flat Friedmann--Lema\^{i}tre--Robertson--Walker universe filled with a (free)
massless scalar field and an arbitrary perfect fluid. For positive energy density in the scalar and fluid, each classical solution has a singularity and expands to infinite volume. When quantising we view the cosmological dynamics in relational terms, using one degree of freedom as a clock for the others. Three natural candidates for this clock are the volume, a time variable conjugate to the perfect fluid, and the scalar field. We have previously shown that requiring unitary evolution in the ``fluid’’ time leads to a boundary condition at the singularity and generic singularity resolution, while in the volume time semiclassical states follow the classical singular trajectories. Here we analyse the third option of using the scalar field as a clock, finding further dramatic differences to the previous cases: the boundary condition arising from unitarity is now at infinity. Rather than singularity resolution, this theory features a quantum recollapse of the universe at large volume, as was shown in a similar context by Paw{\l}owski and Ashtekar. We illustrate the properties of the theory analytically and numerically, showing that the ways in which the different quantum theories do or do not depart from classical behaviour directly arise from demanding unitarity with respect to different clocks. We argue that using a Dirac quantisation would not resolve the issue. Our results further illustrate the problem of time in quantum gravity.
\end{abstract}
 
\section{Introduction}

Time is not a straightforward concept when we are dealing with quantum theories of gravity. The essential reason for this is that general relativity is a generally covariant theory, which can be expressed in an arbitrary coordinate system. There is no notion of time external to the universe; instead we have a dynamically determined metric, which determines how observers locally experience the passing of time. The dynamics of the universe can be described by any globally defined time coordinate, with none of the different coordinates having a preferred status. However, quantum mechanics is based on the Schrödinger equation where `$t$' is an external time parameter that universally defines how time passes. This apparent incompatibility between the paradigm of general relativity and quantum mechanics is known as \emph{the problem of time} \cite{Isham1992,Kuchar2011,Anderson2012}.  

In this work we analyse one of the many aspects of the problem of time in the context of quantum cosmology. Focusing on quantum cosmology rather than a full theory of quantum gravity has many advantages, the main one being perhaps the simplicity of calculations in comparison to full quantum gravity \cite{DeWitt1967}. Quantum cosmological models with only a finite number of degrees of freedom, called \emph{minisuperspace} models \cite{Misner1969}, have been studied for decades to shed light on some of the conceptual and practical open issues of quantum gravity. One of these open issues and the main question we are interested in here is the role of unitarity and self-adjointness for different relational clocks in quantum cosmology. Unitarity is a fundamental assumption in quantum mechanics, made to ensure a consistent probability interpretation, and intimately tied to a particular notion of time evolution. When different notions of time exist, there will be corresponding different and inequivalent notions of unitarity. We will exhibit some of the dramatic consequences of this inequivalence.

There are several strategies to tackle the problem of time. One possibility is to go to a reduced phase space in which one has solved the Hamiltonian constraint for a momentum variable so that the corresponding coordinate can be used as time. This does not solve the problem of time, rather transforms it into the \emph{multiple choice problem} \cite{Kuchar2011}, and the procedure generally gives inequivalent theories for different clock choices (see \cite{Blyth1975} for an early example and \cite{Hajicek2000,*Hajicek2000b,*Malkiewicz2017} for general arguments). Our approach is to first define the quantum dynamics through the Wheeler--DeWitt equation and to then choose an inner product adapted to a particular choice of clock; this will be of Schr\"odinger type if the Wheeler--DeWitt equation is first order in ``time'' and of Klein--Gordon type if it is second order. There is then still an ambiguity of which variable is chosen as a clock. Classically, due to general covariance the viewpoints of different clocks are equivalent and can be translated into each other. However, we have already seen in \cite{Gielen2020} that choosing different variables as clocks leads to inequivalent quantum solutions, in particular regarding singularity resolution. In this work we analyse the problem more deeply by contrasting our previous results with those obtained for a third possible clock choice, and give a more general overview of the inequivalence problem we are facing. 

The universe we study is characterised by a flat FLRW metric and two components of matter, a free massless scalar field $\varphi$ and a perfect fluid. It is possible to write the theory in a Hamiltonian way such that, after a change of coordinates, the Hamiltonian is the same for all choices of perfect fluid. The same dynamics can then be interpreted as corresponding to different types of perfect fluid, i.e.~different equations of state. The perfect fluid is characterised by a conserved quantity. If the fluid is interpreted as describing dark energy, the conserved quantity is the energy density itself, which appears as a conserved momentum conjugate to a time variable $t$. We are then effectively working in \emph{unimodular gravity} \cite{Unruh1989a}. For other interpretations, in which the fluid represents e.g.~dust or radiation, $t$ is not unimodular time but an analogous time coordinate. We have studied the quantisation based on the clock $t$ before; here our choice of relational clock is the scalar field $\varphi$. This clock has been studied previously in very similar models  \cite{Bojowald2010b,Pawlowski2011,Bojowald2016}. We study the resulting quantum theory both analytically and numerically. The Wheeler--DeWitt equation is equivalent to a Klein--Gordon equation on the Rindler wedge with an additional term that corresponds to an attractive or repulsive potential. Demanding that time evolution with respect to $\varphi$ be unitary, we find that for some solutions it is then necessary to impose reflective boundary conditions at $v=\infty$, where the volume of the universe diverges. Due to these boundary conditions, the quantum theory diverges from the classical theory generating a ``quantum recollapse'' (the universe reaches a finite maximum volume where it would classically continue to grow indefinitely), but it does not resolve the singularity as the quantum solutions can follow classical trajectories for small volume. Mathematically the boundary condition we encounter is equivalent to the one arising from self-adjointness of a certain Hamiltonian in an analogous quantum mechanics problem; just as in this case, we find that the subspace of wavefunctions which have unitary dynamics is not unique but depends on a free function, analogous to the usual one-parameter self-adjoint extension problem in quantum mechanics.

Our work extends known results for the quantisation of the same model using different relational clocks. Gryb and Thébault \cite{Gryb2019,Gryb2019b,*Gryb2019c} used the clock $t$ conjugate to the conserved momentum of the perfect fluid and showed that this quantisation has the very attractive property of resolving the big bang singularity. One might argue that this ``fluid'' time, similar to the one arising in unimodular gravity, could be a solution to the problem of time \cite{Unruh1989}, given that it allows for a standard Schr\"odinger quantisation. Interestingly, the requirement of unitarity with respect to evolution in $t$ also leads to a boundary condition, but in this case this leads to singularity resolution rather than a quantum recollapse. In general, we see that the imposition of unitarity is what makes a quantum theory diverge from the classical theory but does not always imply  singularity resolution. Interestingly, one can study the same model using a power of the scale factor as clock, as suggested by Gielen and Turok in \cite{Gielen2017} where the perfect fluid was chosen to represent radiation. For this choice of clock, the quantum theory is unitary without boundary conditions and thus there are no significant divergences from the classical theory for semiclassical states, as we showed in \cite{Gielen2020}. The present work not only adds evidence for the nontriviality of the choice of relational clock, but also explains better why those differences arise in the first place.

In a simpler model \cite{Gotay1984,*Gotay1996} Gotay and Demaret established the conjecture that, if one demands unitary dynamics, ``slow'' clocks resolve the big bang/big crunch singularity, where a clock is slow at a point if it reaches that point at a finite time. Gotay and Demaret only considered  slow clocks at a singularity but we expand their notion to infinity (infinite spatial volume of the universe): we find that slow clocks at infinity trigger a quantum recollapse in the quantum theory. The crucial property of a slow clock is that a classical solution terminates at some point within the range of the clock variable. When building a unitary quantum theory one needs to continue time evolution past this point, resulting in the boundary conditions we observe. This is analogous to the well-known situation in standard quantum mechanics where boundary conditions are needed when a solution terminates in finite time. In  \cite{Gielen2020} we found that the $t$ clock is slow at the singularity, hence forcing singularity resolution when quantising. We also found that the clock associated with the scale factor is fast at the singularity and at infinity, implying that the quantum theory does not need to diverge from the classical theory. In this paper we find that the scalar field clock is slow at infinity. This explains why the three clocks have very different behaviour in the quantum theory.

Rather than choosing particular inner products and hence Hilbert spaces adapted to different clocks, one could use the more covariant approach of Dirac quantisation in order to avoid the multiple choice problem. Here one first defines a kinematical Hilbert space on which constraints are imposed after quantisation. Solutions to the constraints form the physical Hilbert space, on which one can then study dynamics. A canonical way of defining an inner product on the physical Hilbert space is via the so-called group averaging procedure \cite{Ashtekar1995,Marolf2000,*Marolf1995}, which requires the Hamiltonian constraint to be defined as a self-adjoint operator. A recent major result in the setting of Dirac quantisation has been an explicit demonstration, for a general class of systems, that the viewpoints of different clocks are equivalent, and that the quantisation method itself is equivalent to other methods of quantisation of constrained systems \cite{Hoehn2019,*Hoehn2020}. One might hope to use these results to resolve the inequivalence of quantum theories defined with respect to different clocks that we have observed. We will discuss how the methods of Dirac quantisation and group averaging could be applied to our model and show that, in order to separate the Hamiltonian constraint into a canonical form in which it is a sum of a Hamiltonian of the ``clock'' and a Hamiltonian of the ``system'', different clocks require a different choice of lapse function in the definition the Hamiltonian constraint. Hence, even if the constraint (Wheeler--DeWitt) equation is always the same, different clocks require different operators to be self-adjoint on different kinematical Hilbert spaces. The freedom of choosing a lapse in general relativity is of course another aspect of general covariance, which is then seen to be violated even in the setting of Dirac quantisation. New methods are needed to resolve the breaking of general covariance we observe.

The structure of this paper is as follows. In section \ref{classth} we analyse the model classically and derive the Hamiltonian and the relevant Dirac observables. In section \ref{qtumth} we quantise our model using the scalar field as relational clock and construct a normalised basis of allowed wavefunctions. We also discuss the necessary boundary conditions for unitarity. In section \ref{numres} we calculate expectation values of relevant observables numerically for semiclassical states, and show explicitly how the universe recollapses at large volume in situations where it would continue to expand classically. In section \ref{dirac} we compare our clock-dependent quantisation to the framework of Dirac quantisation  before concluding in section \ref{ccl}. \ref{calculations} contains the derivation of some important integrals of Bessel functions that are needed in the main text.

\section{The classical theory}
\label{classth}
In this section we introduce the classical theory whose quantisation we will be studying. 
We are interested in the dynamics of a flat FLRW universe filled with a free massless scalar field $\phi$ and a perfect fluid with general equation of state parameter $w<1$ (where the pressure is $p=w\rho$ in terms of the energy density $\rho$); particularly interesting cases include radiation ($w=\frac{1}{3}$), dust ($w=0$) and dark energy ($w=-1$).
More details on the properties of this model can be found in our previous paper \cite{Gielen2020}. The model has been previously studied in quantum cosmology, e.g.~in \cite{Gryb2019,Gryb2019b,*Gryb2019c,Gielen2017}.

The dynamics of general relativity coupled to a massless scalar and a perfect fluid are defined by an action
\begin{equation}
\fl \mathcal{S} = \int \dd^4 x \left\{\sqrt{-g}\left[\frac{R}{2\kappa}-\frac{1}{2}g^{ab}\partial_a\phi\partial_b\phi -\rho\left(\frac{|J|}{\sqrt{-g}}\right)\right]+J^a(\partial_a\vartheta+\beta_A\partial_a\alpha^A)\right\}\, ,
\end{equation}
where the dynamical variables are the spacetime metric $g_{ab}$, scalar field $\phi$, densitised particle number flux $J^a$ and Lagrange multipliers $\vartheta$, $\beta_A$ and $\alpha^A$. $\rho$, the energy density of the fluid, is a function of $|J|=\sqrt{-g_{ab}J^aJ^b}$ and $\sqrt{-g}$. The action we are using for the perfect fluid is Eq.~(6.10) of \cite{Brown1993}, which describes an isentropic fluid. We have also defined $\kappa=8\pi G$ where $G$ is Newton's constant.

In our model spacetime is a manifold with topology $\mathbb{R}\times\Sigma$ where $\Sigma$ is assumed to be compact (such as a 3-torus); we assume that the matter fields and geometry are homogeneous and (locally) isotropic on each $\Sigma$. We can take the metric to be \begin{equation}
\dd  s^2= -N(\tau)^2\dd \tau^2 + a(\tau)^2 h_{ij}\dd x^i \dd x^j
\label{metric}
\end{equation}
where $h_{ij}$ is a flat metric, $a(\tau)$ is the scale factor and $N(\tau)$ the lapse function. $\phi$ becomes a function of $\tau$ only and the densitised particle number flux must be of the form $J^a= a^3 n\, \delta^a_t$ where $n=n(\tau)$ is the particle number density (see e.g. \cite{Gielen2017} for more details). All Lagrange multipliers are also only functions of $\tau$.

The reduced minisuperspace action, after an integration by parts, is then
\begin{equation}
\mathcal{S} = V_0\int_{\mathbb{R}} \dd \tau\left[-\frac{3\dot{a}^2a}{N\kappa}+\frac{a^3}{2N}\dot\phi^2 -Na^3\rho(n)+a^3 n(\dot\vartheta+\beta_A\dot\alpha^A)\right]\, .
\label{minisuperaction}
\end{equation}
We have implicitly added a boundary term to cancel the boundary contribution from integration by parts, and $V_0=\int_{\Sigma} \dd^3 x\sqrt{h}$ is the coordinate volume of $\Sigma$. The last term in (\ref{minisuperaction}) involving $\beta_A$ and $\alpha^A$ may now be dropped: variation with respect to $\vartheta$ imposes particle number conservation $\frac{\dd}{\dd \tau}(a^3 n) = 0$ and there is no further constraint from these other Lagrange multipliers. The other constraints, requiring the fluid flow to be directed along flow lines labelled by the $\alpha^A$, are trivial in FLRW symmetry.

For a perfect fluid with $p=w\rho$ we have $\rho(n)=\rho_0 n^{1+w}$ for some constant $\rho_0$. Replacing $n$ by another variable $m$ defined by $na^3=(m/\rho_0)^{\frac{1}{1+w}}$ we then finally obtain
\begin{equation}
\mathcal{S} = V_0\int_{\mathbb{R}} \dd \tau\left[-\frac{3\dot{a}^2a}{N\kappa}+\frac{a^3}{2N}\dot\phi^2 -N\frac{m}{a^{3w}}+m \dot\chi\right]
\label{finalaction}
\end{equation}
where we also redefined the Lagrange multiplier to simplify the form of the action (note that the constraint $\frac{\dd}{\dd \tau}(na ^3)=0$ is the same as $\dot{m}=0$). 

The transformation leading to (\ref{finalaction}) is ill-defined for $w=-1$, the case in which the perfect fluid corresponds to dark energy; however, in this case the action (\ref{finalaction}) can be interpreted in terms of unimodular gravity. Unimodular gravity is usually presented as a version of general relativity in which the metric determinant is kept fixed, $\sqrt{-g}=\eta$ in terms of a given volume form $\eta$, which restricts the symmetry group from the full diffeomorphism group to volume-preserving diffeomorphisms (see e.g.~\cite{Unruh1989a}). However, one can go from this restricted formulation to a ``parametrised'' form in which additional fields are introduced which restore the full diffeomorphism symmetry. The action for parametrised unimodular gravity is \cite{Henneaux1989,*Smolin2009}
\begin{equation}
\mathcal{S}_{PUM}=\int \dd^4x \left\{ \frac{\sqrt{-g}}{2\kappa}[ R-2\Lambda]+\Lambda\partial_a T^a\right\}\, .
\label{PUMaction}
\end{equation}
Here $\Lambda$ is a dynamical field; the equation of motion coming from $T^a$ is $\partial_a\Lambda=0$ resulting in the usual statement in unimodular gravity that $\Lambda$ appears as an integration constant, whereas variation with respect to $\Lambda$ fixes $\sqrt{-g}=\kappa\partial_a T^a$. One way of deriving (\ref{PUMaction}) from the usual version of unimodular gravity with $\sqrt{-g}=\eta$ is to promote a set of coordinates $X^A$ in which $\sqrt{-g}=\eta$ holds to fields $X^A(x)$, restoring the full diffeomorphism symmetry in terms of coordinate transformations of the arbitrary  $x$ \cite{Kuchar1991b}. The restriction of (\ref{PUMaction}) to a flat FLRW universe is
\begin{equation}
\mathcal{S}_{PUM} = V_0\int_{\mathbb{R}} \dd \tau\left[-\frac{3\dot{a}^2a}{N\kappa}-Na^3\frac{\Lambda}{\kappa}+\Lambda\dot{T}\right]
\end{equation}
which is exactly of the form (\ref{finalaction}) with $w=-1$ if an additional massless scalar field is coupled. Apart from this case of a ``dynamical'' dark energy which provided the motivation for \cite{Gryb2019,Gryb2019b,*Gryb2019c}, other cases of interest are $w=\frac{1}{3}$ which corresponds to radiation, as studied in \cite{Gielen2017,Gielen2016}, and dust ($w=0$) which is often added as a matter component in quantum cosmology \cite{Gotay1984,*Gotay1996, Ali2018, *Husain2019}, typically because it allows for deparametrisation.  We will shortly see explicitly that deparametrisation is possible for any choice of parameter $w$ by choosing an appropriate gauge.

\subsection{Hamiltonian formulation}

We will now show that the Hamiltonian resulting from the canonical analysis of the action (\ref{finalaction}) takes the same form for any value of $w$ as long as $w<1$. We find the canonical momenta
\begin{equation}
\pi_a = -V_0\frac{6\dot{a}a}{N\kappa}\,,\quad \pi_\phi=V_0\frac{a^3}{N}\dot\phi
\label{momentadef}
\end{equation}
and $m$ is the conjugate momentum to $\chi$, $\{\chi,m\}=1$. The Hamiltonian is
\begin{equation}
\mathcal{H} = N \left[  -\frac{1}{12}\frac{\kappa \pi_a^2}{V_0 a}+\frac{1}{2}\frac{\pi_\phi^2}{V_0 a^3}+V_0\frac{m}{a^{3w}}\right]\,.
\label{hamilt1}
\end{equation}
In the following we set $\kappa=1$ to simplify the notation. To then bring (\ref{hamilt1}) into a common form for all values of $w$, we can apply the canonical transformation
\begin{equation}
v=4\sqrt{\frac{V_0}{3}}\,\frac{a^{\frac{3(1-w)}{2}}}{1-w},\quad \pi_v=\sqrt{\frac{1}{12V_0}}\,\pi_a\,a^{\frac{3w-1}{2}}
\label{transfo}
\end{equation}
and rescale the scalar field variables by $\varphi = \sqrt{\frac{3}{8}}(1-w)\phi,\;\pi_\varphi = \sqrt{\frac{8}{3}}\frac{\pi_\phi}{1-w}
$ to obtain
\begin{equation}
\mathcal{H} = \tilde{N} \left[-\pi_v^2+\frac{\pi_\varphi^2}{v^2}+\lambda\right]
\label{Hamiltonian}
\end{equation}
where we also defined $\lambda=V_0m$ and a new lapse $\tilde{N}=Na^{-3w}=N\left(\frac{16V_0}{3v^2(1-w)^2}\right)^{\frac{w}{1-w}}$. In this form $w$ no longer appears explicitly.

We have hence shown that (\ref{Hamiltonian}) defines the Hamiltonian for a flat FLRW cosmology coupled to a free massless scalar field and a perfect fluid with arbitrary equation of state parameter $w<1$. Since the scale factor $a$ is generally taken to be positive (and we will also assume in the following that $v\ge 0$), the appearance of fractional powers in (\ref{transfo}) is not problematic. Nevertheless, for particular values of $w$ the variables $v$ and $\pi_v$ have a more direct geometrical interpretation. In particular for $w=-1$, the case of dark energy, we have
\begin{equation}
v=2\sqrt{\frac{V_0}{3}}a^3,\quad\pi_v=\sqrt{\frac{1}{12V_0}}\,\frac{\pi_a}{a^2}
\end{equation}
and $v$ is proportional to the volume of space while $\pi_v$ is proportional to the Hubble rate $\frac{\dot{a}}{aN}$ (cf.~(\ref{momentadef})). Having this particular case in mind and being consistent with the notation of our previous paper \cite{Gielen2020} suggested the notation $v$ for the variable defined in (\ref{transfo}) although for $w\neq -1$ this variable would not be proportional to a volume.

The theory we consider is reparametrisation invariant, as is evident by the fact that the lapse $N$ or $\tilde{N}$ can be chosen arbitrarily. Nevertheless, the theory takes a particularly simple form if one chooses $\tilde{N}=1$: in this case the Hamiltonian (\ref{Hamiltonian}) can be written as $\mathcal{H}=-\mathcal{H}_0(v,\pi_v,\pi_\varphi)+\lambda$ with $\mathcal{H}_0=\pi_v^2-\frac{\pi_\varphi^2}{v^2}$ and the Hamiltonian constraint becomes $\mathcal{H}_0 = \lambda$: $\lambda$ then plays the r\^{o}le of the energy of the system defined by $v$ and $\varphi$ and their conjugate momenta. Notice that in such a gauge $\frac{\dd t}{\dd \tau}=1$, where $t=\frac{\chi}{V_0}$ is conjugate to $\lambda$; $t$ becomes the time variable for the evolution of $v$ and $\varphi$, which justifies the notation. For dark energy with $w=-1$, this preferred choice corresponds to $N= a^{-3}$, a unimodular gauge in which the metric has constant determinant and time is proportional to the four-volume of the universe \cite{Brown1989}. For radiation it is conformal time $N=a$. All other choices of perfect fluid similarly have a preferred time coordinate in which the dynamics take the simplest possible form. This time variable is the one used in deparametrisation where the reparametrisation invariance of the theory is gauge-fixed, e.g.~in the commonly studied case where the perfect fluid represents (nonrotating) dust \cite{Brown1994,Husain2011}.\footnote{An interesting extension of the framework studied here, in which a single perfect fluid is present, is to study models with multiple fluids which each provide a possible notion of time \cite{Magueijo2021}.}

We can write the Hamiltonian (\ref{Hamiltonian}) in the form
\begin{equation}
\mathcal{H} = \tilde{N} \left[g^{AB}\pi_A \pi_B+\lambda\right]
\label{metricformham}
\end{equation}
where $\pi_A=(\pi_v,\pi_\varphi)$ and $g^{AB}=\left(\matrix{ -1 & 0 \cr 0 & \frac{1}{v^2}}\right)$. This notation shows that the Hamiltonian is equivalent to the one of a relativistic particle moving in the 2-dimensional manifold parametrised by $v$ and $\varphi$ and with metric $g_{AB}$, the inverse of $g^{AB}$, where $\lambda$ is analogous to the squared mass. In the quantisation that we mainly study in this paper,  $\varphi$ is defined to be the time coordinate; the space parametrised by $v$ and $\varphi$ is then equivalent to the Rindler wedge in two-dimensional Minkowski spacetime. Its boundary is the big bang/big crunch singularity $v=0$. Note also that in this analogy $\lambda>0$ would correspond to a ``tachyon'' with negative mass squared whereas $\lambda<0$ would be a massive particle.

\subsection{Classical solutions}
\label{classsol}

The Hamiltonian equations can be solved explicitly to obtain the classical solutions. Since $\lambda$ is a conserved quantity, solutions can be classified by whether $\lambda$ is positive, negative or zero. We saw above that in the case where one thinks of a perfect fluid with $w=-1$, $\lambda$ is essentially the cosmological constant and thus could take either sign. For other types of perfect fluid one might assume that particle number density and energy density must be positive and only consider $\lambda>0$. The classical solutions we present are always well-defined for any interpretation of the perfect fluid matter.

It is insightful to write the classical solutions in relational form, i.e.~to express some phase space variables as functions of the others. When written in this form the classical solutions provide the starting point for the construction of Dirac observables which we will present in the next section; in this sense, solutions expressed in relational terms directly correspond to observables.

The momenta $\lambda$ and $\pi_\varphi$ are constants of motion. Among the remaining four canonical variables, $t$, $v$ and $\varphi$ are all possible candidates for relational clocks\footnote{The canonical momentum $\pi_v$ would also be a good clock everywhere: its equation of motion gives $\frac{\dd\pi_v}{\dd \tau}=2\tilde{N}\frac{\pi_\varphi^2}{v^3}>0$. A similar ``extrinsic'' clock was studied in a slightly simpler model in \cite{Blyth1975}.}. In particular, as we stated above, in the gauge $\tilde{N}=1$ we have $\frac{\dd t}{\dd \tau}=1$ and $t$ becomes the evolution parameter for the other variables. In any gauge $\frac{\dd t}{\dd \tau}=\tilde{N}>0$ and hence $t$ is always monotonic; we can always express all other variables as functions of $t$.

We presented some of these solutions in \cite{Gielen2020} but repeat them here to keep the presentation self-contained. For $\lambda\neq 0$ and $\pi_\varphi\neq 0$ we have
\begin{equation}
v(t)=\sqrt{-\frac{\pi_\varphi^2}{\lambda}+4\lambda(t-t_0)^2}\,,\quad \varphi(t)=\frac{1}{2}\log\left|\frac{\pi_\varphi-2\lambda(t-t_0)}{\pi_\varphi+2\lambda(t-t_0)}\right|+\varphi_0\, ,
\label{scale-factor}
\end{equation}
where $t_0$ and $\varphi_0$ are integration constants. The big bang/big crunch singularity $v=0$ is reached for the two solutions of $|t-t_0|=\frac{|\pi_\varphi|}{2|\lambda|}$; for $\lambda>0$ the solution has two branches, a contracting one ending in a big crunch and an expanding one emerging from the big bang, whereas for $\lambda<0$ the solution is defined in between the two singular points, expanding and then recollapsing. In this last case $v$ would not be a globally defined clock due to the recollapse of the universe. The scalar field $\varphi$ diverges logarithmically (in $t$) at the singularities.

For $\lambda=0$ the solutions take the slightly different form 
\begin{equation}
v(t)=2\sqrt{|\pi_\varphi||t-t_0|}\,,\quad \varphi(t)=\frac{1}{2}{\rm sgn}(\pi_\varphi(t-t_0))\log\left|\frac{t}{t_0}-1\right|+\varphi_0\,.
\label{lambda0ttime}
\end{equation}

Since we will later use $\varphi$ as a clock, it is instructive to express the other dynamical variables as functions of $\varphi$ as well. Notice that $\varphi$ is always a globally defined clock as long as $\pi_\varphi\neq 0$, which we assume throughout. If $\varphi$ is used as clock we need to distinguish between positive and negative $\lambda$ solutions. Namely, for $\lambda>0$ we find
\begin{equation}
v(\varphi)=\frac{|\pi_\varphi|}{\sqrt{\lambda}\left|\sinh(\varphi-\varphi_0)\right|}\,,\quad t(\varphi)=-\frac{\pi_\varphi}{2\lambda}\coth(\varphi-\varphi_0)+t_0
\label{fcts-of-phi}
\end{equation}
whereas for $\lambda<0$
\begin{equation}
v(\varphi)=\frac{|\pi_\varphi|}{\sqrt{-\lambda}\cosh(\varphi-\varphi_0)}\,,\quad t(\varphi)=-\frac{\pi_\varphi}{2\lambda}\tanh(\varphi-\varphi_0)+t_0\,.
\end{equation}
The case $\lambda=0$ is conceptually rather different if the scalar field $\varphi$ is used as a clock. In this case only, $\varphi$ takes all values from $-\infty$ to $\infty$ in each of the two (expanding and contracting) branches of the solution, and can only parametrise one half of the full solution (\ref{lambda0ttime}). This is because for $\lambda=0$ the field $\varphi$ keeps growing logarithmically at large volume, unlike for $\lambda>0$ where it approaches the constant value $\varphi=\varphi_0$. Hence for $\lambda=0$ one needs to choose which branch of the solution one is in. From the equation of motion for $v$\,,
\begin{equation}
\frac{\dd v}{\dd \tau} = \{v,\mathcal{H}\} = -2\tilde{N}\pi_v\,,
\end{equation}
one sees that since $\tilde{N}>0$ the expanding (contracting) branch corresponds to negative (positive) $\pi_v$. In particular this implies that if $\lambda\ge 0$ the sign of $\pi_v$ does not change during the evolution. With this in mind we can then give the two possible solutions with $\lambda=0$ as
\begin{equation}
\fl v(\varphi)=2\sqrt{|\pi_\varphi\,t_0|}\,e^{-{\rm sgn}(\pi_\varphi\pi_v)(\varphi-\varphi_0)}\,,\quad t(\varphi)=t_0-|t_0|\,{\rm sgn}(\pi_v)e^{-2\, {\rm sgn}(\pi_\varphi\pi_v)(\varphi-\varphi_0)}\,.
\label{varphisolutions}
\end{equation}

\begin{figure}
\begin{center}
\includegraphics[scale=0.8]{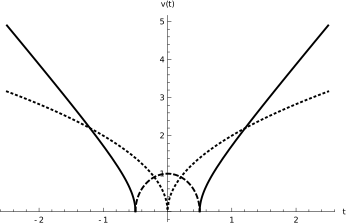}
\includegraphics[scale=0.8]{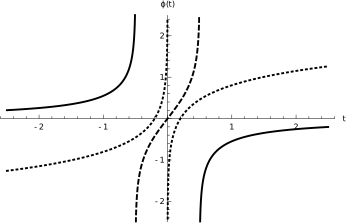}
\caption{Classical solutions $v(t)$ and $\varphi(t)$ for $\pi_\varphi=1$ and $\lambda=1$ (solid curve), $\lambda=-1$ (dashed curve) and $\lambda=0$ (dotted curve). At the big bang/big crunch singularities $t=\pm\frac{1}{2}$ (solid and dashed) and $t=0$ (dotted) the scalar field $\varphi$ diverges logarithmically.}
\label{fig1}
\end{center}
\end{figure}

\begin{figure}
\begin{center}
\includegraphics[scale=0.8]{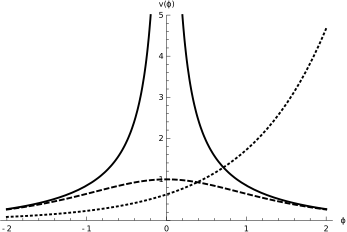}
\includegraphics[scale=0.8]{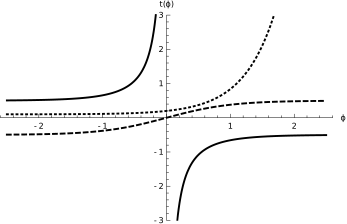}
\caption{Classical solutions $v(\varphi)$ and $t(\varphi)$ for $\pi_\varphi=1$ and $\lambda=1$ (solid curve), $\lambda=-1$ (dashed curve) and $\lambda=0$ (dotted curve). Notice that $v=\infty$ is reached at $\varphi=0$ for $\lambda> 0$ but at $\varphi=+\infty$ for $\lambda=0$, where we have chosen the expanding branch of the solution. Here $\varphi=-\infty$ is always the big bang; for $\lambda\neq 0$, $\varphi=+\infty$ is the big crunch.}
\label{fig2}
\end{center}
\end{figure}

These classical solutions are plotted in Figures \ref{fig1} and \ref{fig2}. In these plots we use $\lambda=\pm 1$ for the positive and negative $\lambda$ cases and have set the arbitrary integration constants $\varphi_0$ and $t_0$ to zero where possible. (The quantities $\lambda$ and $\pi_\varphi$ are really dimensionful, so implicit in this is a choice of units for $V_0$ in (\ref{transfo}) in addition to setting $\kappa=8\pi G=1$.)

Figures \ref{fig1} and \ref{fig2} also illustrate that using $v$ as a time parameter for the other variables requires a choice of expanding or contracting branch similar to the one for the $\varphi$ clock when $\lambda=0$. All classical solutions with $\pi_\varphi\neq 0$ have a contracting and an expanding part; $v$ can only parametrise one of these. For $\lambda\ge 0$ the contracting and expanding branches are not connected, and one can view $v$ as a global clock on one of the branches. For $\lambda<0$, there is a recollapse within a single branch beyond which $v$ fails to be a good clock. In the quantum theory, both branches are present as a superposition within a generic quantum state, which makes the $v$ time hard to interpret. In \cite{Gielen2020} we studied a quantum theory that uses $v$ as a clock, and focused on semiclassical states peaked on the expanding branch for $\lambda>0$, for which $v$ does define a good (global) clock.

\subsection{Dirac observables}

For any theory with local gauge symmetry, observables are required to be gauge-invariant. In case of diffeomorphism symmetry this rather restricts the possibilities for local observables: thinking of these as defined at a certain ``time'', since time is just a coordinate label a local observable must be a constant of motion (see e.g.~\cite{Unruh1989}). This apparent absence of any notion of time evolution appears puzzling until one realises that in a diffeomorphism-invariant theory dynamics must be expressed in relational terms, as the change in one phase space variable with respect to another \cite{Rovelli1990,*Rovelli1991,Dittrich2006,*Dittrich2007,Tambornino2011}. Rather than relying on coordinate labels, a {\em relational} reference frame can be defined in terms of suitable matter fields, usually taken to be scalars \cite{Brown1994,DeWitt1962,*Marolf1994,*Ferrero2020}. In a spatially homogeneous model the only relational ``coordinate'' required is time. Here we define a few (Dirac) observables in the classical theory that will become important when we analyse the physical content of a quantum theory based on choosing $\varphi$ as the clock. These observables extend those given for other choices of time variable given in our previous work \cite{Gielen2020}.

The Hamiltonian \eqref{Hamiltonian} is of the form $\mathcal{H}=\tilde{N} \C$ where $\tilde{N}$ is a Lagrangian multiplier and $\C$ a function of the phase space variables, usually referred as a totally constrained Hamiltonian. This form is a common feature of all systems with diffeomorphism invariance \cite{GaugeSystems}. The presence of the Lagrange multiplier $\tilde{N}$ forces the relation
\begin{equation}
\C=-\pi_v^2+\frac{\pi_\varphi^2}{v^2}+\lambda\approx 0,
\label{ham-cons}
\end{equation}
called the Hamiltonian constraint. 
\eqref{ham-cons} defines the constraint surface $\mathscr{C}$, where the use of the sign $\approx$ means that a relation holds only in $\mathscr{C}$ and not in the entire phase space. Observables must be invariant under the gauge transformations (time reparametrisations) generated by $\C$; hence on $\mathscr{C}$ all Dirac observables $\mathcal{O}$ must satisfy
\begin{equation}
\lbrace \mathcal{C},\mathcal{O} \rbrace \approx 0 \iff -\frac{2\pi_\varphi^2}{v^3}\pdv{\mathcal{O}}{\pi_v}-\pdv{\mathcal{O}}{t}-\frac{2\pi_\varphi}{v^2}\pdv{\mathcal{O}}{\varphi}+2\pi_v\pdv{\mathcal{O}}{v} \approx 0\,.
\label{poisson-braket}
\end{equation}
We can immediately see that \eqref{poisson-braket} implies that $\lbrace\H,\mathcal{O}\rbrace\approx 0$, which confirms that all Dirac observables are constants of motion.  For example, any function of the conserved quantities $\lambda$ and $\pi_\varphi$ is a Dirac observable. However, these observables are frozen, i.e.~they do not reflect any evolution of the system.

By defining observables that parametrise the evolution of different phase space variables with respect to given reference variable, we can describe evolution with respect to a specific clock. In our case, as we will use $\varphi$ as a clock, we are interested in the relational observables $v(\varphi=\varphi_1)$ and $t(\varphi=\varphi_1)$, which give the values of $v$ and $t$ when the clock $\varphi$ takes the value $\varphi_1$. Although these observables are constants of motion for a given $\varphi_1$, by letting $\varphi_1$ vary we obtain a set of \emph{complete observables}. 

Fix a point $P=(t_{{\rm i}},v_{{\rm i}},\varphi_{{\rm i}},\lambda,\pi_{v_{{\rm i}}},\pi_\varphi)\in \mathscr{C}$. The Dirac observables $v(\varphi=\varphi_1)$ and $t(\varphi=\varphi_1)$ then give the values of $v$ and $t$ when $\varphi=\varphi_1$ for the solution which originates from the initial data $P$. Concretely, we find that these observables are given by
\begin{equation}
\fl
v(\varphi=\varphi_1)=\left\lbrace \begin{array}{ll}
\displaystyle{\frac{\abs{\pi_\varphi}}{\sqrt{\lambda}\abs{\sinh\left(\varphi_1-\varphi_{{\rm i}}+\sgn{\pi_{v_{{\rm i}}}}{\rm arsinh}\left(\frac{\pi_\varphi}{v_{{\rm i}} \sqrt{\lambda}}\right)\right)}}}\,, & \lambda>0 \\
\displaystyle{\frac{\abs{\pi_\varphi}}{\sqrt{-\lambda}\,\cosh\left(\varphi_1-\varphi_{{\rm i}}+\sgn{\pi_\varphi\pi_{v_{{\rm i}}}}{\rm arcosh}\left(\frac{\abs{\pi_\varphi}}{v_{{\rm i}} \sqrt{-\lambda}}\right)\right)}}\,, & \lambda<0 \\
\displaystyle{v_{{\rm i}} e^{-\sgn{\pi_{\varphi}\pi_{v_{{\rm i}}}}(\varphi_1-\varphi_{{\rm i}})}}\,, & \lambda =0
\end{array} \right.
\label{v-phi1}
\end{equation}
and 
\begin{equation}
\fl
t(\varphi=\varphi_1)=\left\lbrace \begin{array}{ll}
\displaystyle{-\frac{\pi_\varphi}{2\lambda}\coth\left(\varphi_1-\varphi_{{\rm i}}+{\rm artanh}\left( \frac{\pi_\varphi}{\pi_{v_{{\rm i}}} v_{{\rm i}}} \right) \right)}+t_{{\rm i}}+\frac{ \pi_{v_{{\rm i}}}v_{{\rm i}}}{2\lambda}\,, & \lambda>0 \\
\displaystyle{-\frac{\pi_\varphi}{2\lambda}\tanh\left(\varphi_1-\varphi_{{\rm i}}+{\rm artanh}\left( \frac{\pi_{v_{{\rm i}}}v_{{\rm i}}}{\pi_\varphi} \right) \right)+t_{{\rm i}}+\frac{\pi_{v_{{\rm i}}}v_{{\rm i}}}{2\lambda}}\,, & \lambda<0 \\
 \displaystyle{\left(t_{{\rm i}}+\frac{v_{{\rm i}}}{4\pi_{v_{{\rm i}}}}\right)\left(1-\frac{v_{{\rm i}}}{4t_{{\rm i}}\pi_{v_{{\rm i}}}+v_{{\rm i}}}e^{-2\,\sgn{\pi_\varphi\pi_{v_{{\rm i}}}}(\varphi_1-\varphi_{{\rm i}})}\right)}\,, & \lambda=0\,.
\end{array}\right.
\label{t-phi1}
\end{equation}
By varying $P$ these observables extend to all of $\mathscr{C}$.
These observables differ from the simpler types of observables built from constants of motion $\lambda $ and $\pi_\varphi$ in that their Poisson bracket vanishes only in the constraint surface $\mathscr{C}$, since $\lbrace \C, v(\varphi=\varphi_1) \rbrace=f \C\neq 0$ for some non trivial function $f$, whereas we have $\lbrace\C,\lambda\rbrace=\lbrace \C ,\pi_\varphi\rbrace=0$ everywhere in the phase space. This property is common for relational Dirac observables of the type we have defined here.

What these observables describe is the (future and past) behaviour of the classical solution determined by specifying an initial data point $P$. We can give explicit expressions for the observables since we know all classical solutions explicitly.

\section{Quantum theories based on different clocks}
\label{qtumth}
This section deals with the quantisation of the cosmological model introduced above, extending our previous work in \cite{Gielen2020}.
The Hamiltonian constraint \eqref{ham-cons} becomes the Wheeler--DeWitt equation of the quantum theory. The Wheeler--DeWitt equation can be solved explicitly in terms of Bessel functions. We introduce an inner product and Hilbert space based on using the scalar field $\varphi$ as a clock. Demanding unitarity in the sense of conservation of the inner product in $\varphi$ then imposes constraints on the allowed wavefunctions. We compare the resulting quantum theory to the theories discussed in \cite{Gielen2020} where the role of clock was played by the ``Schrödinger clock'' $t$ or a volume variable.

Recall from \eqref{metricformham} that the Hamiltonian constraint can be written as
\begin{equation}
\C=-\pi_v^2+\frac{\pi_\varphi^2}{v^2}+\lambda = g^{AB}\pi_A\pi_B+\lambda
\end{equation}
where $g^{AB}=\left(\matrix{ -1 & 0 \cr 0 & \frac{1}{v^2}}\right)$ is the inverse metric of the Rindler wedge parametrised by $v$ and $\varphi$. This rewriting suggests a natural choice for the operator ordering in the Wheeler--DeWitt equation for the quantum theory: we replace the term $g^{AB}\pi_A\pi_B$ with $-\hbar^2\square$ where
\begin{equation}
\square= \frac{1}{\sqrt{-g}}\frac{\partial}{\partial q^A}\left(g^{AB}\sqrt{-g}\frac{\partial}{\partial q^B}\right)=-\frac{1}{v}\frac{\partial}{\partial v}\left(v\frac{\partial}{\partial v}\right)+\frac{1}{v^2}\frac{\partial^2}{\partial\varphi^2}
\label{beltrami}
\end{equation}
is the Laplace--Beltrami operator on the Rindler wedge. The resulting Wheeler--DeWitt equation is
\begin{equation}
\left( \hbar^2\frac{\partial^2}{\partial v^2}+\frac{\hbar^2}{v}\pdv{}{v}-\frac{\hbar^2}{v^2}\frac{\partial^2}{\partial\varphi^2}-\im\hbar \pdv{}{t}\right)\Psi(v,\varphi,t)=0.
\label{wdw}
\end{equation}
There is always an ordering ambiguity when writing the Wheeler--DeWitt equation. One approach to dealing with this ambiguity is to include free parameters into the Wheeler--DeWitt equation which correspond to different orderings. One can then study the impact of these parameters on the theory (see e.g.~\cite{Steigl2005}). Here we instead follow the perspective advocated by Hawking and Page \cite{Hawking1985} that there is a preferred ordering for the operators which makes the Wheeler--DeWitt equation covariant under coordinate transformations on the Rindler wedge parametrised by $(v,\varphi)$. This choice corresponds to \eqref{wdw} and is unique up to the addition of  a term $\hbar^2 \xi R$ where $R$ is the Ricci scalar of minisuperspace \cite{DeWitt1957, *Moss1988, Halliwell1988}. In our case, as the Rindler wedge is flat, there is a unique covariant ordering prescription.

The Wheeler--DeWitt equation \eqref{wdw} can be solved by separation of variables using the ansatz $\Psi(v,\varphi,t)=\nu(\varphi)\psi(v)e^{\im\lambda \frac{t}{\hbar}}$ which leads to the two equations
\begin{equation}
\frac{\nu''(\varphi)}{\nu(\varphi)}= A
\label{phi}
\end{equation}
and 
\begin{equation}
v^2\psi''(v)+v\psi'(v)+\left(\frac{\lambda}{\hbar^2}v^2- A\right)\psi(v)=0 \,.
\label{bessel}
\end{equation}
\eqref{phi} is straightforward to solve; if we assume $A\neq 0$ then depending on the sign of $A$ the solutions are either real exponentials $e^{\kappa\varphi}$ or imaginary exponentials $e^{\im k\varphi}$. \eqref{bessel} is well-known as Bessel's equation; again depending on the sign of $A$ its solutions are either real order Bessel functions $J_{\pm\abs{\kappa}}\left(\frac{\sqrt{\lambda}}{\hbar}v\right)$ or purely imaginary order Bessel functions $J_{\pm\im\abs{k}}\left(\frac{\sqrt{\lambda}}{\hbar}v\right)$.  The general solution to the Wheeler--DeWitt equation can then be written as
\begin{eqnarray}
\fl
\Psi(v,\varphi,t)&=&\int_{-\infty}^{\infty} \frac{\dd \lambda}{2\pi \hbar}\int_{-\infty}^{\infty}\frac{\dd k}{2\pi} e^{\im k\varphi}e^{\im\lambda \frac{t}{\hbar}}\left[ \alpha(k,\lambda)J_{\im\abs{k}}\left(\frac{\sqrt{\lambda}}{\hbar}v\right) +\beta(k,\lambda)J_{-\im\abs{k}}\left(\frac{\sqrt{\lambda}}{\hbar}v\right)\right] \nonumber\\ \fl
 &+&\int_{-\infty}^{\infty} \frac{\dd \lambda}{2\pi \hbar}\int_{-\infty}^{\infty}\frac{\dd \kappa}{2\pi} e^{ \kappa\varphi}e^{\im\lambda \frac{t}{\hbar}}\left[ \gamma(\kappa,\lambda)J_{\abs{\kappa}}\left(\frac{\sqrt{\lambda}}{\hbar}v\right) +\epsilon(\kappa,\lambda)J_{-\abs{\kappa}}\left(\frac{\sqrt{\lambda}}{\hbar}v\right)\right]
\label{sols}
\end{eqnarray}
where at this point $\alpha$, $\beta$, $\gamma$ and $\epsilon$ are arbitrary complex functions. In the case where $\lambda$ is negative, we use the convention
\begin{equation}
\J{x}{\lambda}=J_{x}\left(\im\frac{\sqrt{-\lambda}}{\hbar}v\right)=e^{\frac{\im x\pi}{2}}I_{x}\left(\frac{\sqrt{-\lambda}}{\hbar}v\right)
\label{analytic-continuation}
\end{equation}
for any (real or imaginary) $x$, where $I_x$ denotes the modified Bessel function of the first kind. 

\subsection{Scalar field $\varphi$ as a relational clock}
\label{phi-clock}

Up to this point the Wheeler--DeWitt equation \eqref{wdw} is simply a differential equation in all the arguments of $\Psi$, and there is no notion of time evolution. To give an interpretation and a physical meaning to the solutions \eqref{sols}, we choose an internal variable to serve as relational clock and build an inner product (and then a Hilbert space) with respect to the remaining variables. Time evolution then corresponds to defining states or observables as functions of the possible values taken by the clock variable. The clock, despite being an internal variable of the system, is then treated similarly to the external time parameter in quantum mechanics: it has no uncertainty. Due to the general covariance of general relativity there is (at least locally) no a priori preferred clock variable; some clocks may have the advantage of being monotonic and hence globally defined. 

In this paper, we introduce a quantum theory based on the field $\varphi$ as a clock, given that $\varphi$ is always monotonic as we saw in section \ref{classsol}. We then compare this theory to theories based on using $t$ or $v$ as a clock; these have been studied in detail in \cite{Gielen2020}, but we will briefly review them to give a meaningful comparative between the three quantisations studied.

The Wheeler--DeWitt equation \eqref{wdw} can be rewritten as
\begin{equation}
\left(\hbar^2\frac{\partial^2}{\partial\varphi^2} -\hbar^2\left(\pdv{}{\log(v/v_0)}\right)^2+\im \hbar v^2\pdv{}{t}\right)\Psi(v,\varphi,t)=0,
\label{wdw2}
\end{equation}
where $v_0$ is a parameter of dimensionality length$^{3/2}$  needed for dimensional reasons. From this point of view, the equation can be regarded as a Klein--Gordon-type equation in the variables $\log(v/v_0)$ and $\varphi$ plus a third term that, after making the ansatz $\Psi(v,\varphi,t)= \psi(v,\varphi)e^{\im\lambda \frac{t}{\hbar}}$, is similar to a potential depending on $v$ and $\lambda$. In \cite{Gielen2020} we used this observation as a starting point for a theory built on the clock $\log(v/v_0)$; here instead we propose the inner product
\begin{equation}
\fl
\braket{\Psi}{\Phi}_\varphi=\im\int_{-\infty}^{\infty} \dd t \int_{0}^{\infty} \frac{\dd v}{v}\left( \bar{\Psi}(v,\varphi,t)\pdv{}{\varphi}\Phi(v,\varphi,t)-\Phi(v,\varphi,t)\pdv{}{\varphi}\bar{\Psi}(v,\varphi,t)\right).
\label{phi-inner-prod}
\end{equation}
Note that we integrate over $t$ and $v$ but not over $\varphi$, which is the clock of the theory. The subindex $\varphi$ reminds us that the inner product is in principle a function of $\varphi$, in contrast to inner products defined with respect to other clocks. The differential operator appearing in (\ref{phi-inner-prod}) is $n^\mu \partial_\mu\equiv \frac{1}{v}\partial_\varphi$ where $n^\mu$ is the normal to $\varphi={\rm const}$ surfaces in the Rindler wedge metric \eqref{metricformham}.

We now demand unitarity, i.e.~require that $\pdv{}{\varphi}\braket{\Psi}{\Phi}_{\varphi}=0$ for any $\Psi$ and $\Phi$, in order to be able to have a meaningful (conventional) probability interpretation of this quantum theory. This results in the boundary condition
\begin{equation}
\int \dd t  \left[ v\bar{\Psi}\pdv{}{v}\Phi-v\Phi\pdv{}{v}\bar{\Psi}\right]_{v=0}^{v=\infty}=0.
\label{bound}
\end{equation}
This condition is {\em not} automatically satisfied by all solutions of (\ref{wdw2}); instead it imposes a restriction on the general solutions \eqref{sols}. 

Using again the ansatz $\Psi(v,\varphi,t)= \psi(v,\varphi)e^{\im\lambda \frac{t}{\hbar}}$, one can see that \eqref{bound} is equivalent to the condition arising from demanding self-adjointness of the Hamiltonian
\begin{equation}
\hat{\mathfrak{H}}=-\hbar^2\frac{\partial^2}{\partial u^2}-\lambda e^{2u}
\label{fr-ham}
\end{equation}
with respect to a standard $L^2$ inner product on the real line $\mathbb{R}$ parametrised by the coordinate $u$. For $\lambda>0$ this Hamiltonian contains an attractive potential in which classically a particle could reach infinity ($u=\infty$ or $v=\infty$) in a finite time. In the quantum theory with such a potential one then finds that the Hamiltonian $\hat{\mathfrak{H}}$ is not self-adjoint on $L^2(\mathbb{R})$ and one needs to impose a boundary condition which amounts to a reflection of the wavefunction from $v=\infty$. There is a one-parameter family of linear subspaces of states that satisfy \eqref{bound} at infinity. The characterisation of the Hamiltonian and the derivation of this family are presented in a very comprehensive way in \cite{Fredenhagen2003} in the context of S-branes; these results were first derived in \cite{Kobayashi1996, *Fulop1995}. In our case, since $\lambda$ is not a fixed parameter but a dynamical variable the one-parameter freedom of choosing a subspace of states becomes a choice of free function of $\lambda$. We will rederive the solutions to the boundary condition \eqref{bound}, reproducing the results found in the literature. However, the normalisation of our wavefunctions is not the same due to the fact that we are working with a Klein--Gordon, not a Schr\"odinger inner product. For $\lambda<0$ the potential in \eqref{fr-ham} is repulsive; in this case the Hamiltonian is already self-adjoint without a boundary condition. The solutions for this case are well-known, see e.g.~\cite{DHoker1982}.

An additional issue in our discussion is that the inner product \eqref{phi-inner-prod} is not positive definite, as is usually the case for Klein--Gordon-type inner products that include derivatives. An inner product that is not positive definite can also not be used for a consistent (Born) probability interpretation. Our approach to this issue is to first impose unitarity, i.e.~obtain an inner product that is conserved over time. This means we restrict our solutions to those satisfying the boundary condition (\ref{bound}). The resulting subspace of wavefunctions splits into mutually orthogonal subspaces of respectively positive, negative or zero norm under \eqref{phi-inner-prod}. One can then redefine the inner product by changing the overall sign on the negative norm sector. The explicit construction of a  positive definite inner product over the space of all solutions to the Wheeler--DeWitt equation is not needed as solutions that do not satisfy (\ref{bound}) do not have a physical interpretation in our setting: no meaningful probability distribution can be associated to a state whose norm is not conserved over time. 

In the rest of this section we give a detailed derivation of the form of wavefunctions that are normalisable in the inner product \eqref{phi-inner-prod} while also being compatible with \eqref{bound}. For each separate positive value of $\lambda$ the general solution to the boundary condition \eqref{bound} is the one required to ensure self-adjointness of the Hamiltonian \eqref{fr-ham} with the same value of $\lambda$; requiring normalisability puts further restrictions on states. Our results are also compatible with the analysis of \cite{Pawlowski2011} where a quantum theory in $\varphi$ time was constructed for a cosmological model with massless scalar field and again a fixed $\lambda>0$. The authors of \cite{Pawlowski2011} define a Dirac quantisation on a kinematical Hilbert space rather than fixing a clock variable before quantisation, and also do not use the $v$ variable but a dual representation in terms of a variable $b$ corresponding to the Hubble parameter. Hence the results are not directly comparable but the physical and mathematical features found there mirror exactly those found in our case. In section \ref{dirac} we will comment on the connection of our quantum theory with the framework of Dirac quantisation, and argue that the required boundary conditions should indeed be the same in both frameworks.

Readers not interested in the details of the derivation can find the general form of a normalisable state whose norm is preserved under $\varphi$ evolution in \eqref{general-solution}. 

\subsection{Normalisability and boundary conditions for $\lambda>0$}

We first consider a general state with $\gamma(\kappa,\lambda)=\epsilon(\kappa,\lambda)=0$ in (\ref{sols}), 
\begin{equation}
\fl
\Psi_1(v,\varphi,t)=\int_{0}^{\infty} \frac{\dd \lambda}{2\pi \hbar}\int_{-\infty}^{\infty}\frac{\dd k}{2\pi} e^{\im k\varphi}e^{\im\lambda \frac{t}{\hbar}}\left[ \alpha(k,\lambda)J_{\im\abs{k}}\left(\frac{\sqrt{\lambda}}{\hbar}v\right) +\beta(k,\lambda)J_{-\im\abs{k}}\left(\frac{\sqrt{\lambda}}{\hbar}v\right)\right]\,.
\label{generalwf}
\end{equation}
This is to separately discuss the cases of Bessel functions of real and imaginary order, which have very different asymptotic behaviour. For (\ref{generalwf}) we then find
\begin{eqnarray}
\fl\braket{\Psi_1}{\Psi_1}_\varphi &=-\int \frac{\dd k_1 \dd k_2}{(2\pi)^2}\,\frac{\dd \lambda}{2\pi \hbar}\, \frac{\dd v}{v}\, e^{\im (k_2-k_1)\varphi}(k_1+k_2)\times \\
\fl&  \left[ \bar{\alpha}_1\alpha_2\J{-\im \abs{k_1}}{\lambda}\J{\im \abs{k_2}}{\lambda} +  \bar{\alpha}_1\beta_2\J{-\im \abs{k_1}}{\lambda}\J{-\im \abs{k_2}}{\lambda} \right. \nonumber \\
\fl&+  \left.\bar{\beta}_1\alpha_2\J{\im \abs{k_1}}{\lambda}\J{\im \abs{k_2}}{\lambda} + \bar{\beta}_1\beta_2 \J{\im \abs{k_1}}{\lambda}\J{-\im \abs{k_2}}{\lambda} \right] \nonumber
\end{eqnarray}
where we use the abbreviation $\alpha_i$ for $\alpha(k_i,\lambda)$ and $\beta_i$ for $\beta(k_i,\lambda)$. Here and in the following we need to make use of explicit expressions for the integral of products of two Bessel functions over $v$, which are derived in \ref{calculations}. Here the relevant integral is \eqref{2-Bessel-int-1}.  After regrouping terms, we find 
\begin{eqnarray}
\fl
\braket{\Psi_1}{\Psi_1}_\varphi&=&  - \int\frac{\dd k_1 \dd k_2}{(2\pi)^2}\,\frac{\dd \lambda}{2\pi \hbar} \, e^{\im (k_2-k_1)\varphi}(k_1+k_2)\times 
\label{longpsiintegral}
 \\
\fl & & \left\{{\rm PV}\frac{2\,\im}{\pi(k_1^2-k_2^2)}\left(\sinh\left((\abs{k_1}+\abs{k_2})\frac{\pi}{2}\right)\left[\bar{\alpha}_1\alpha_2-\bar{\beta}_1\beta_2\right]\right.\right.\nonumber
\\\fl & & \qquad\left.\left.+\sinh\left((\abs{k_1}-\abs{k_2})\frac{\pi}{2}\right)\left[\bar{\alpha}_1\beta_2-\bar{\beta}_1\alpha_2\right]\right)\right.\nonumber \\
\fl &&\quad+ 2\,\frac{\sinh((\abs{k_1}+\abs{k_2})\frac{\pi}{2})}{\abs{k_1}+\abs{k_2}}\delta(\abs{k_1}-\abs{k_2})\left[ \bar{\alpha}_1\alpha_2+\bar{\beta_1}\beta_2 \right]\nonumber \\
\fl &&\quad+ \left. 2\,\frac{\sinh((\abs{k_1}-\abs{k_2})\frac{\pi}{2})}{\abs{k_1}-\abs{k_2}}\delta(\abs{k_1}+\abs{k_2})\left[ \bar{\alpha}_1\beta_2+\bar{\beta}_1\alpha_2 \right]\right\}\nonumber
\end{eqnarray}
where ${\rm PV}$ denotes the Cauchy principal value, i.e.~a definition of the integral in terms of a symmetric limit around the singular point $k_1=k_2$.

We can see that the last line does not contribute to the final result. We can simplify the other Dirac delta term using
\begin{equation}
\fl(k_1+k_2)\delta(\abs{k_1}-\abs{k_2})=(k_1+k_2)\left[\delta(k_1+k_2)+\delta(k_1-k_2) \right]=2k_1 \delta(k_1-k_2)\,.
\label{delta}
\end{equation}
Since the delta function $\delta(\abs{k_1}-\abs{k_2})$ forces $k_1=k_2$, the contribution coming from this term is independent of $\varphi$. The other two terms cannot be further simplified and hence, in order for the inner product to be independent of $\varphi$, they must vanish. This leads to the condition
\begin{equation}
Y(k_1,k_2,\lambda,\varphi)-Y(2k_2-k_1,k_2,\lambda,\varphi)=0
\label{cond}
\end{equation}
to be satisfied for all $k_1, k_2,\lambda$ and $\varphi$, where
\begin{eqnarray}
Y(k_1,k_2,\lambda,\varphi)&=&e^{\im (k_2-k_1)\varphi}\left(\sinh\left((\abs{k_1}+\abs{k_2})\frac{\pi}{2}\right)\left[\bar{\alpha}_1\alpha_2-\bar{\beta}_1\beta_2\right]\right.\nonumber
\\
&&\left.+\sinh\left((\abs{k_1}-\abs{k_2})\frac{\pi}{2}\right)\left[\bar{\alpha}_1\beta_2-\bar{\beta}_1\alpha_2\right]\right)
\end{eqnarray}
and we have used the fact that the principal value integral only depends on the part of $Y$ that is odd with respect to reflection around the singular point $k_1=k_2$. We can now expand the condition (\ref{cond}) to linear order around $k_1=k_2$ where it vanishes, finding
\begin{eqnarray}
\fl
\alpha(k,\lambda)\bar\beta(k,\lambda)-\bar\alpha(k,\lambda)\beta(k,\lambda)-\cosh(k\pi)\left(|\alpha(k,\lambda)|^2-|\beta(k,\lambda)|^2\right)&&
\\
\fl
+\frac{2}{\pi}\sinh(k\pi)\left(\beta(k,\lambda)\frac{\partial\bar\beta(k,\lambda)}{\partial k}-\alpha(k,\lambda)\frac{\partial\bar\alpha(k,\lambda)}{\partial k}+\im\varphi(|\alpha(k,\lambda)|^2-|\beta(k,\lambda)|^2)\right)&=&0\,.\nonumber
\end{eqnarray}
Only the last term inside the large brackets depends on $\varphi$ so we must have
\begin{equation}
|\alpha(k,\lambda)|^2 = |\beta(k,\lambda)|^2\qquad \implies \;e^{\im \chi(k,\lambda)}\beta(k,\lambda)=\alpha(k,\lambda)
\label{firstcond}
\end{equation}
where $\chi(k,\lambda)$ is a function taking values in $[-\pi,\pi)$. The remaining terms then also vanish if
\begin{equation}
\pi \sin(\chi(k,\lambda))+\sinh(k\pi)\pdv{}{k}\chi(k,\lambda)=0\,.
\end{equation}
The general solution to this equation can be written as
\begin{equation}
\chi(k,\lambda)=-2\arctan\left[\theta(\lambda)\coth\left(\frac{\abs{k}\pi}{2}\right)\right]
\label{phase}
\end{equation}
in terms of a free function $\theta(\lambda)$. There would also be the freedom to multiply $\chi$ by an overall sign which can be different for positive and negative $k$ (noting that $\chi(k,\lambda)$ is ill-defined at $k=0$) but this is fixed by demanding that \eqref{phase} is a solution to \eqref{cond}, hence ensuring that we have found the general solution to the boundary condition \eqref{bound}. 

Two obvious solutions to \eqref{cond} are $\beta(k,\lambda)=\pm\alpha(k,\lambda)$ for some (or all) $\lambda$. The ``$+$'' solution corresponds to $\chi(k,\lambda)=\theta(\lambda)=0$ for these values of $\lambda$ but the ``$-$'' solution is $\chi(k,\lambda)=-\pi$ which formally corresponds to $\theta(\lambda)=\infty$, which must hence be included as a possible choice. The free function $\theta(\lambda)$ taking values in $\mathbb{R}\cup\{\infty\}$ is then the analogue of the self-adjoint extension parameter for the Hamiltonian \eqref{fr-ham} discussed in \cite{Fredenhagen2003,Kobayashi1996, *Fulop1995}. 

For any state of the form \eqref{generalwf} with \eqref{firstcond} and \eqref{phase} we then find
\begin{equation}
\braket{\Psi_1}{\Psi_1}_\varphi = -\int_{-\infty}^{\infty}\dk{k} \int_{0}^{\infty} \dl{\lambda}\,\frac{2\sinh(k\pi)}{\pi}\abs{\alpha(k,\lambda)}^2\,. 
\label{newinnerprod}
\end{equation}
We see that $k>0$ modes have a negative contribution to the norm, whereas $k<0$ modes have positive norm. One possible approach would now be to only include $k<0$ (positive frequency) modes into the subspace of allowed wavefunctions, and indeed in the later numerical analysis we will restrict to this subspace. However, it is also possible to redefine the inner product so that it becomes positive definite for all $k$, and all these modes can be considered as physical states. This is what we did in \cite{Gielen2020} when considering a volume variable as time; an inner product that is positive definite for all $k$ would also come out of a group averaging construction, which we will compare with below.

Since the $k>0$ and $k<0$ modes are decoupled after imposing \eqref{cond}, we can define such a positive definite inner product by
\begin{equation}
\braket{\Psi_1}{\Psi_1}_{\varphi'}= \braket{\Psi_1}{\Psi_1}_{\varphi, k<0}-\braket{\Psi_1}{\Psi_1}_{\varphi,k>0}
\label{inner-prod-mod}
\end{equation} 
where the subindices $k<0$ and $k>0$ refer to the values of $k$ in the integration in \eqref{newinnerprod}. 
Explicitly, for a general solution to the boundary condition \eqref{bound} we define
\begin{equation}
\braket{\Psi_1}{\Psi_1}_{\varphi'}=\int_{-\infty}^{\infty} \dk{k}\int_{0}^{\infty}\dl{\lambda} \frac{2\sinh(\abs{k}\pi)}{\pi}\abs{\alpha(k,\lambda)}^2
\end{equation}
which is now manifestly positive definite. Our constructions also ensure that this inner product is conserved under evolution in $\varphi$, i.e.~time evolution is unitary for these states.

A normalised solution to the Wheeler--DeWitt equation satisfying the boundary condition and built only out of oscillatory modes (i.e.~with $\gamma(\kappa,\lambda)=\epsilon(\kappa,\lambda)=0$ in \eqref{generalwf}) can then be written as
\begin{eqnarray}
\fl
\Psi_1(v,\varphi,t)=\int_{-\infty}^{\infty} \dk{k} \int_{0}^{\infty}\dl{\lambda} e^{\im k\varphi}e^{\im \frac{\lambda}{\hbar}t} \alpha(k,\lambda) \sqrt{\frac{2\pi}{\sinh(\abs{k}\pi)}}\Re\left[e^{\im \frac{ \chi(k,\lambda)}{2}}\J{\im |k|}{\lambda}  \right] 
\label{im-k-wave-fct-1}
\end{eqnarray}
where $\int_{-\infty}^\infty \dk{k}\int_{0}^\infty \dl{\lambda} \abs{\alpha(k,\lambda)}^2=1$ and $\Re$ denotes the real part. Here
\begin{equation}
e^{\im \frac{ \chi(k,\lambda)}{2}}= e^{-\im \arctan\left[\theta(\lambda)\coth\left(\frac{\abs{k}\pi}{2}\right)\right]} = \sqrt{\frac{\sinh\left(\frac{\abs{k}\pi}{2}\right)-\im \theta(\lambda)\cosh\left(\frac{\abs{k}\pi}{2}\right)}{\sinh\left(\frac{\abs{k}\pi}{2}\right) +\im \theta(\lambda)\cosh\left(\frac{\abs{k}\pi}{2}\right)}}
\end{equation}
after rewriting the arctangent in terms of a logarithm.

One can slightly simplify this final expression by introducing a function $\kappa_0$ with
\begin{equation}
\theta(\lambda)=\tan\left(\kappa_0(\lambda)\frac{\pi}{2}\right)
\label{kappa0}
\end{equation}
where $\kappa_0$ takes values in the interval $[0,2)$. In this new notation, $\kappa_0(\lambda)=1$ corresponds to the possible choice $\theta(\lambda)=\infty$ discussed earlier. Using \eqref{kappa0} we can rewrite \eqref{im-k-wave-fct-1} as
\begin{eqnarray}
\Psi_1(v,\varphi,t)&=&\int_{-\infty}^\infty \dk{k}\int_0^{\infty}\dl{\lambda} e^{\im k\varphi}e^{\im \frac{\lambda}{\hbar}t}\alpha(k,\lambda)\sqrt{\frac{2\pi}{\sinh(\abs{k}\pi)}}\times \label{im-k-wave-fct-2} \\ &&\Re\left[\sqrt{\frac{\sinh((\abs{k}-\im \kappa_0(\lambda))\frac{\pi}{2})}{\sinh((\abs{k}+\im\kappa_0(\lambda))\frac{\pi}{2})}}\J{\im |k|}{\lambda} \right], \nonumber
\end{eqnarray}
This form for the allowed wavefunctions corresponds to the one given in \cite{Fredenhagen2003}, but the normalisation is different due to a different choice of inner product here.

Now we consider solutions formed of real exponentials in $\varphi$. That is, in \eqref{sols} we set $\alpha(k,\lambda)=\beta(k,\lambda)=0$ and consider a wavefunction 
\begin{equation}
\fl
\Psi_2(v,\varphi,t)=\int_0^{\infty} \dl{\lambda} \int_{-\infty}^{\infty} \dk{\kappa} e^{\kappa \varphi} e^{\im \lambda \frac{t}{\hbar}}\left[ \gamma(\kappa,\lambda)\J{\abs{\kappa}}{\lambda}+\epsilon(\kappa,\lambda)\J{-\abs{\kappa}}{\lambda} \right].
\end{equation}
The inner product \eqref{phi-inner-prod} of such a state with itself is 
\begin{eqnarray}
\fl
\braket{\Psi_2}{\Psi_2}_\varphi &=&\im \int \frac{\dd \kappa_1 \dd \kappa_2}{(2\pi)^2}\frac{\dd \lambda}{2\pi \hbar} \frac{\dd v}{v} e^{ (\kappa_1+\kappa_2)\varphi}(\kappa_2-\kappa_1)\times \nonumber \\
\fl &&  \left[ \bar{\gamma}_1\gamma_2\J{ \abs{\kappa_1}}{\lambda}\J{\abs{\kappa_2}}{\lambda} \right. + \bar{\gamma}_1\epsilon_2\J{ \abs{\kappa_1}}{\lambda}\J{- \abs{\kappa_2}}{\lambda}  \\
\fl &&  +\bar{\epsilon}_1\gamma_2\J{ -\abs{\kappa_1}}{\lambda}\J{ \abs{\kappa_2}}{\lambda}\left. + \bar{\epsilon}_1\epsilon_2 \J{-\abs{\kappa_1}}{\lambda}\J{- \abs{\kappa_2}}{\lambda} \right] \nonumber
\end{eqnarray}
where we use again the abbreviation $\gamma_i=\gamma(\kappa_i,\lambda)$ and $\epsilon_i=\epsilon(\kappa_i,\lambda)$. In the Appendix we show that the integral over $v$ can only be defined (even in a distributional sense) when the sum of the orders of the Bessel functions is strictly positive, in which case it is given by \eqref{2-Bessel-int-2}. We must hence assume $\epsilon(\kappa,\lambda)=0$ to get a normalisable state. We then find
\begin{equation}
\fl
\braket{\Psi_2}{\Psi_2}_{\varphi}=- \frac{2\,\im}{\pi}\int \frac{\dd \kappa_1 \dd\kappa_2}{(2\pi)^2}\dl{\lambda} e^{(\kappa_1+\kappa_2)\varphi}\frac{\sin((\abs{\kappa_1}-\abs{\kappa_2})\frac{\pi}{2})}{\kappa_1+\kappa_2}\bar\gamma(\kappa_1,\lambda)\gamma(\kappa_2,\lambda)\,.
\end{equation} 
The only way to ensure $\pdv{}{\varphi}\braket{\Psi_2}{\Psi_2}_\varphi=0$ is to set
\begin{equation}
\sin\left((\abs{\kappa_1}-\abs{\kappa_2})\frac{\pi}{2}\right)=0\quad\implies\quad  \abs{\kappa_1}-\abs{\kappa_2}=2n\,, \ n\in \bZ\,.
\end{equation}
This condition can be solved separately for each value of $\lambda$; but given a fixed $\lambda$, only a discrete set of values for $\kappa$ is allowed, namely those satisfying
\begin{equation}
\abs{\kappa}=\kappa_0'(\lambda) +2n \ \mathrm{for} \ \mathrm{some} \ n\in\bN_0
\label{kappa0'}
\end{equation}
where $\kappa_0'(\lambda)$ is an arbitrary function of $\lambda$ which we can choose to take values in $[0,2)$. 
The general form of such $\Psi_2$ whose norm is preserved under evolution in $\varphi$ is then
\begin{equation}
\fl
\Psi_2 =\int_{0}^{\infty} \dl{\lambda} e^{\im \frac{\lambda}{\hbar} t} \left[\sum_{n=0}^{\infty} \left( \gamma_n^+(\lambda) e^{(\kappa_0'(\lambda)+2n)\varphi}+ \gamma_n^-(\lambda) e^{-(\kappa_0'(\lambda)+2n)\varphi}\right) \J{\kappa_0'(\lambda)+2n}{\lambda}\right]\,.
\label{kappas}
\end{equation} 
Such a wavefunction has zero norm, which is indeed $\varphi$ independent. These states do not have a classical analogue; one would have to interpret them as configurations for which $\pi_\varphi^2<0$ in \eqref{ham-cons}, analogous to tunnelling solutions under a potential barrier in standard quantum mechanics. In quantum cosmology, similar states which ``decay'' in relational time have been discussed as ``quantum puff'' universes by Misner \cite{Misner1972}. The interpretation of such states is in general far from clear if one demands a unitary quantum theory, as e.g.~mentioned in Blyth's PhD thesis \cite{Blyth1974}.
In our case they have norm zero, so they have no influence in the probabilistic interpretation of the theory.

Now that we have analysed the two sectors $k$ and $\kappa$ separately, we require that the inner product $\braket{\Psi_1}{\Psi_2}_\varphi$ is independent of $\varphi$ for two normalisable states $\Psi_1$ and $\Psi_2$ given respectively by \eqref{im-k-wave-fct-1} and \eqref{kappas}. We find
\begin{eqnarray}
\fl
\braket{\Psi_1}{\Psi_2}_{\varphi}&=&\im \int \dk{k} \frac{\dd \lambda}{2\pi\hbar}\frac{\dd v}{v}\sum_{n=0}^\infty \left( e^{(\kappa_0'(\lambda)+2n-\im k)\varphi}\bar{\alpha}(k,\lambda)\gamma_n^+(\lambda)(\kappa_0'(\lambda)+2n+\im k)  \right.\nonumber  \\
\fl&&\quad\left.+e^{(-\kappa_0'(\lambda)-2n-\im k)\varphi}\bar{\alpha}(k,\lambda)\gamma_n^-(\lambda)(-\kappa_0'(\lambda)-2n+\im k)\right)\times\nonumber
\\\fl && \sqrt{\frac{2\pi}{\sinh(\abs{k}\pi)}}\Re\left[e^{\im \frac{\chi(k,\lambda)}{2}}\J{\kappa_0'(\lambda)+2n}{\lambda}\J{\im \abs{k}}{\lambda}  \right]\,.
\end{eqnarray}
We can then use the result \eqref{2-Bessel-int-3} to compute the $v$ integral to obtain
\begin{eqnarray}
\fl
\braket{\Psi_1}{\Psi_2}_{\varphi} &=&\frac{2\,\im}{\pi}\int \dk{k}\dl{\lambda}\sum_{n=0}^\infty \left( e^{(\kappa_0'(\lambda)+2n-\im k)\varphi}\frac{\bar{\alpha}(k,\lambda)\gamma_n^+(\lambda)}{\kappa_0'(\lambda)+2n-\im k}  -e^{(-\kappa_0'(\lambda)-2n-\im k)\varphi}\frac{\bar{\alpha}(k,\lambda)\gamma_n^-(\lambda)}{\kappa_0'(\lambda)+2n+\im k}\right)\nonumber
\\\fl && \times\sqrt{\frac{2\pi}{\sinh(\abs{k}\pi)}}\Re\left[e^{\im \frac{\chi(k,\lambda)}{2}}\sin\left((\kappa_0'(\lambda)+2n-\im \abs{k})\frac{\pi}{2}\right) \right] .
\end{eqnarray}
To ensure $\pdv{}{\varphi} \braket{\Psi_1}{\Psi_2}_\varphi=0$ we must demand that $\Re\left[ e^{\im \frac{\chi(k,\lambda)}{2}}\sin\left((\kappa_0'(\lambda)+2n-\im \abs{k})\frac{\pi}{2}\right)\right]$ vanishes, or equivalently that
\begin{equation}
\tan\left(\kappa_0'(\lambda)\frac{\pi}{2}\right)=-\tan\left(\frac{\chi(k,\lambda)}{2}\right)\tanh\left(\frac{\abs{k}\pi}{2}\right)\,.
\end{equation}
Using \eqref{phase} and \eqref{kappa0} this condition becomes
\begin{eqnarray}
\tan\left(\kappa_0'(\lambda)\frac{\pi}{2}\right)=\tan\left(\kappa_0(\lambda)\frac{\pi}{2}\right)\,,
\end{eqnarray}
so that if we choose $\kappa_0(\lambda)=\kappa_0'(\lambda)$ for all $\lambda$ the condition $\pdv{}{\varphi}\braket{\Psi_1}{\Psi_2}_\varphi=0$ is satisfied. 

In conclusion, the most general normalised wavefunction built only from $\lambda>0$ modes and which satisfies the boundary condition \eqref{bound} is
\begin{eqnarray}
\fl
\Psi(v,\varphi,t)&=&\int_{-\infty}^\infty \dk{k}\int_{0}^{\infty}\dl{\lambda} e^{\im k\varphi}e^{\im \frac{\lambda}{\hbar}t}\alpha(k,\lambda)\sqrt{\frac{2\pi}{\sinh(\abs{k}\pi)}}\times\nonumber
\\\fl&&\qquad\Re\left[\sqrt{\frac{\sinh((\abs{k}-\im \kappa_0(\lambda))\frac{\pi}{2})}{\sinh((\abs{k}+\im\kappa_0(\lambda))\frac{\pi}{2})}}\J{\im \abs{k}}{\lambda} \right]  \\
\fl
&&+\int_{0}^{\infty} \dl{\lambda} e^{\im \frac{\lambda}{\hbar} t} \left[\sum_{n=0}^{\infty} \left( \gamma_n^+(\lambda) e^{(\kappa_0(\lambda)+2n)\varphi}+ \gamma_n^-(\lambda) e^{-(\kappa_0(\lambda)+2n)\varphi}\right) \J{\kappa_0(\lambda)+2n}{\lambda}\right]\nonumber
\label{wf-phi}
\end{eqnarray}
where $\int_{-\infty}^{\infty} \dk{k} \int_{0}^{\infty} \dl{\lambda}\abs{\alpha(k,\lambda)}^2=1$. Notice that there is no restriction on the values of $\gamma_n^\pm$ since the real Bessel modes do no contribute to the norm of this state. For both real and imaginary order Bessel functions, we saw that for each mode either the boundary condition \eqref{bound} or the requirement of normalisability means that out of the two independent solutions to the Wheeler--DeWitt equation only one is allowed.

\subsection{State space for $\lambda<0$ and summary}
\label{phi-summary}

We now focus on states built from solutions to the Wheeler--DeWitt
equation for which $\lambda<0$. The behaviour of classical solutions is
very different for positive and negative $\lambda$; for $\lambda<0$ they
recollapse at some maximum volume rather than accelerating to infinity
as for $\lambda>0$. We will of course see this difference reflected in
the quantum theory: there is no need for a boundary condition at $v=\infty$ as in the positive $\lambda$ case (recall that the corresponding quantum mechanics Hamiltonian \eqref{fr-ham} is already self-adjoint if $\lambda<0$).

As in the previous discussion, we first restrict to Bessel functions of imaginary order. Consider the wavefunction
\begin{equation}
\fl
\Psi_3(v,\varphi,t)=\int_{-\infty}^{0} \frac{\dd \lambda}{2\pi
\hbar}\int_{-\infty}^{\infty}\frac{\dd k}{2\pi} e^{\im
k\varphi}e^{\im\lambda \frac{t}{\hbar}}\left[
\zeta(k,\lambda)I_{\im\abs{k}}\left(\frac{\sqrt{-\lambda}}{\hbar}v\right) +\xi(k,\lambda)I_{-\im\abs{k}}\left(\frac{\sqrt{-\lambda}}{\hbar}v\right)\right]
\label{negativelambda}
\end{equation}
where we use the convention (\ref{analytic-continuation}) to define the
Bessel functions with imaginary argument and we have absorbed the factor
$e^{\pm\frac{\abs{k}\pi}{2}}$ into the functions $\zeta(k,\lambda)$ and
$\xi(k,\lambda)$.

As for the case $\lambda>0$, we must require that these states are
normalisable and their norm is preserved under time evolution.
Normalisability poses an immediate problem: the large $v$ asymptotic
behaviour of the modified Bessel functions of the first kind is
\begin{equation}
I_x\left(\frac{\sqrt{-\lambda}}{\hbar}v\right) \sim
\sqrt{\frac{\hbar}{2\pi
v\sqrt{-\lambda}}}\,e^{\frac{\sqrt{-\lambda}}{\hbar}v}
\end{equation}
and hence generic states of the form (\ref{negativelambda}) cannot be
normalisable in the inner product (\ref{phi-inner-prod}) since the
integral over $v$ is badly divergent.

In fact the asymptotic form of the modified Bessel functions of the
first kind for large argument contains both an exponentially growing and
an exponentially decaying part. The growing part is even under
$x\rightarrow -x$. This motivates the definition of a different Bessel
function, the modified Bessel function of the second kind
\begin{equation}
K_x\left(\frac{\sqrt{-\lambda}}{\hbar}v\right) =
\frac{\pi}{2}\,\frac{I_x\left(\frac{\sqrt{-\lambda}}{\hbar}v\right)-I_{-x}\left(\frac{\sqrt{-\lambda}}{\hbar}v\right)}{\sin(x\pi)}
\end{equation}
which decays exponentially at large $v$. These modified Bessel functions
correspond to the usual normalisable quantum-mechanical wavefunctions in
a classically forbidden region, such as the large $v$ region when
$\lambda<0$; indeed these are the solutions usually considered for a Schr\"odinger equation with repulsive, exponential potential \cite{DHoker1982}.

We hence restrict ourselves to wavefunctions of the form
\begin{equation}
\Psi_3(v,\varphi,t)=\int_{-\infty}^0 \dl{\lambda}\int_{-\infty}^\infty \dk{k}e^{\im k\varphi}e^{\im \frac{\lambda}{\hbar}t} \eta(k,\lambda)\K{\im \abs{k}}{-\lambda}\,.
\end{equation}
Notice that modified Bessel functions of the second
kind are always real for real and positive argument. 
For the inner product \eqref{phi-inner-prod} of such a state we then find
\begin{eqnarray}
\fl
\braket{\Psi_3}{\Psi_3}_\varphi=-\int \dl{\lambda}\frac{\dd k_1 \dd k_2}{(2\pi)^2}\frac{\dd v}{v}(k_1+k_2) e^{\im(k_2-k_1)\varphi}\bar{\eta}_1\eta_2 \K{\im \abs{k_1}}{-\lambda}\K{\im \abs{k_2}}{-\lambda}
\end{eqnarray}
where we are using the usual shorthand notation $\eta_i=\eta(k_i,\lambda)$. The integral over $v$ can be calculated using \eqref{kintegral}, resulting in
\begin{equation}
\fl
\braket{\Psi_3}{\Psi_3}_\varphi = -\int \dl{\lambda}\frac{\dd k_1 \dd k_2}{(2\pi)^2}e^{\im(k_2-k_1)\varphi}\frac{\pi^2(k_1+k_2) \bar{\eta}_1\eta_2}{2\abs{k_1}\sinh(\abs{k_1}\pi)}\left[\delta(\abs{k_1}-\abs{k_2})+\delta(\abs{k_1}+\abs{k_2})\right] \,.
\end{equation}
Note that there is no contribution to the $v$ integral from the upper limit $v=\infty$, which is another way of seeing that the boundary condition \eqref{bound} is already satisfied here.

 Once again, the factor $\delta(\abs{k_1}+\abs{k_2})$ does not contribute to the integral. We can also use \eqref{delta} to write $(k_1+k_2)\delta(\abs{k_1}-\abs{k_2})=2k_1\delta(k_1-k_2)$. We obtain
\begin{equation}
\braket{\Psi_3}{\Psi_3}_\varphi=-\frac{\pi}{2}\int \dl{\lambda}\dk{k} \,\frac{\abs{\eta(k,\lambda)}^2}{\sinh(k\pi)}\,.
\end{equation}
We see that the positive $k$ modes have a negative contribution to the norm, again due to the fact that our inner product is of Klein--Gordon form and not positive definite. Hence, given that the positive and negative $k$ modes are decoupled in the inner product, we can define a new inner product for these modes by
\begin{equation}
\braket{\Psi_3}{\Psi_3}_{\varphi'}=\braket{\Psi_3}{\Psi_3}_{\varphi,k<0}-\braket{\Psi_3}{\Psi_3}_{\varphi,k>0}
\end{equation}
where again, $k>0$ and $k<0$ refer to the values of $k$ when integrating. Thus, with this new inner product the states have a squared norm
\begin{equation}
\braket{\Psi_3}{\Psi_3}_{\varphi'}=\frac{\pi}{2}\int_{-\infty}^0 \dl{\lambda}\int_{-\infty}^{\infty}\dk{k}\ \frac{\abs{\eta(k,\lambda)}^2}{\sinh(\abs{k}\pi)}\geq 0.
\end{equation}

Finally, we need to consider real exponential solutions for $\lambda<0$. As a candidate for a normalisable state one could define
\begin{equation}
\Psi_4(v,\varphi,t)=\int_{-\infty}^0 \dl{\lambda} \int_{-\infty}^{\infty}\dk{\kappa} e^{\kappa\varphi}e^{\im \frac{\lambda}{\hbar}t}\omega(\kappa,\lambda) \K{\kappa}{-\lambda}\, .
\end{equation}
However, by looking at \eqref{2-KBessels-int-2} we see that there are no values of $\kappa$ that make $\braket{\Psi_4}{\Psi_4}$ converge, hence such states are not allowed in the theory. 

In conclusion we can now give the most general normalised solution to the Wheeler--DeWitt equation which solves the boundary condition \eqref{bound}:
\begin{eqnarray}
\fl
\Psi(v,\varphi,t)&=&\int_{-\infty}^\infty \dk{k}\int_{0}^{\infty}\dl{\lambda} e^{\im k\varphi}e^{\im \frac{\lambda}{\hbar}t}\alpha(k,\lambda)\sqrt{\frac{2\pi}{\sinh(\abs{k}\pi)}}\times\nonumber
\\\fl&&\qquad\Re\left[\sqrt{\frac{\sinh(\frac{\pi}{2}(\abs{k}-\im \kappa_0(\lambda)))}{\sinh(\frac{\pi}{2}(\abs{k}+\im\kappa_0(\lambda)))}}\J{\im \abs{k}}{\lambda} \right] \nonumber  \\
\fl
&&+\int_{0}^{\infty} \dl{\lambda} e^{\im \frac{\lambda}{\hbar} t} \left[\sum_{n=0}^{\infty} \left( \gamma_n^+(\lambda) e^{(\kappa_0(\lambda)+2n)\varphi}+ \gamma_n^-(\lambda) e^{-(\kappa_0(\lambda)+2n)\varphi}\right) \J{\kappa_0(\lambda)+2n}{\lambda}\right]\nonumber
\\
\fl
&& + \int_{-\infty}^\infty \dk{k}\int_{-\infty}^{0}\dl{\lambda} \e^{\im k\varphi}e^{\im \frac{\lambda}{\hbar}t} \sqrt{\frac{2\sinh(\abs{k}\pi)}{\pi}}\eta(k,\lambda)\K{i\abs{k}}{-\lambda}
\label{general-solution}
\end{eqnarray} 
where $\int_0^\infty \dl{\lambda}\int_{-\infty}^{\infty} \dk{k} \abs{\alpha(k,\lambda)}^2+\int_{-\infty}^{0} \dl{\lambda} \int_{-\infty}^{\infty} \dk{k} \abs{\eta(k,\lambda)}^2=1$. As we discussed before, there is a freedom in defining this solution space which is manifest in the free choice of the function $\kappa_0(\lambda)$ taking values in $[0,2)$. This freedom corresponds to the one-parameter freedom in choosing a self-adjoint extension for a Hamiltonian in standard quantum mechanics, as we discussed around \eqref{fr-ham}. Here we are effectively dealing with an independent self-adjoint extension problem for each value of $\lambda$, given that modes for different $\lambda$ are decoupled in our inner product, and so the freedom is now the choice of an arbitrary function of $\lambda$. One could ask what features of the quantum theory are sensitive to this choice; this question is discussed in the closely related formalism in \cite{Pawlowski2011} where it is shown that the impact of choosing different parameters (for a given fixed value of $\lambda$) is essentially negligible in the behaviour of relevant observables. In the numerical analysis of certain semiclassical states below we will make a particular choice.

Most importantly perhaps for the physical interpretation of the theory, we found that modes with $\lambda>0$, which correspond to classical solutions which can reach infinity in finite time, must satisfy a boundary condition which amounts to these modes being reflected from $v=\infty$. This is to ensure that the quantum theory remains unitary where the classical theory terminates. (Recall from section \ref{classth} that for $\lambda>0$ the universe reaches infinite volume in finite at a finite value of $\varphi$.) Below we will see explicitly that this reflecting boundary condition implies that, rather than reaching infinite volume, solutions turn around when reaching a finite maximal volume.

\subsection{Comparison with other choices of relational clock}
\label{comparsec}

We have already discussed that the model we study contains several good candidates for a clock variable. In our previous work \cite{Gielen2020}, we studied two different relational clocks, $t$ and $\log(v/v_0)$. (Recall from section \ref{classth} that $t$ is the variable conjugate to $\lambda$, and has the interpretation of unimodular time when $\lambda$ is interpreted as dark energy.)
Here we summarise the results of this previous work (referring to \cite{Gielen2020} for details) to compare them to the theory studied in this paper, where $\varphi$ is the clock. In section \ref{numres} we will extend the comparison to a numerical analysis.

We can regard the Wheeler--DeWitt equation \eqref{wdw} as a Schrödinger equation in $t$ with Hamiltonian $\hat{\mathcal{H}}_s$ where
\begin{equation}
\fl
\im \hbar\pdv{}{t} \Psi(v,\varphi,t)=-\hat{\mathcal{H}}_s\Psi(v,\varphi,t)\, , \hspace{4mm} \hat{\mathcal{H}}_s=\hbar^2 \left( -\frac{\partial^2}{\partial v^2}-\frac{1}{v}\pdv{}{v}+\frac{1}{v^2}\frac{\partial^2}{\partial\varphi^2} \right)\, .
\label{Hs}
\end{equation} 
This suggests an interpretation of the quantum theory as describing evolution in $t$, and the definition of an Schr\"odinger inner product
\begin{equation}
\braket{\Psi}{\Phi}_t =\int_0^\infty \dd v \int_{-\infty}^\infty \dd \varphi \ v \, \bar{\Psi}(v,\varphi,t)\Phi(v,\varphi,t)\, .
\end{equation}
Note that because the Wheeler--DeWitt equation only contains a first derivative in $t$, there are no time derivatives in the inner product. The Hilbert space of this theory is $L^2(\mathcal{R}, \dd v \dd \varphi \sqrt{-g})$ where $\mathcal{R}$ is the Rindler wedge and $\sqrt{-g}=v$ for the  metric \eqref{metricformham}. This inner product is positive definite. 

 The Wheeler--DeWitt equation can be transformed into a Schrödinger equation with a radial $1/r^2$ potential: with the ansatz $\Psi(v,\varphi,t)=v^{-1/2}\omega(v)e^{\im k\varphi}e^{\im\lambda\frac{t}{\hbar}}$ \eqref{Hs} becomes 
\begin{equation}
-\hbar^2 \frac{\partial^2}{\partial v^2}\omega(v)-\hbar^2 \frac{k^2+\frac{1}{4}}{v^2}\omega(v)=\lambda \omega(v)\, .
\end{equation}
Properties of such a potential in the Schr\"odinger equation have been analysed in \cite{Narnhofer1974,*Kunstatter2009}. In general one distinguishes several cases depending on the (dimensionless) strength of the potential. In our case, this coefficient is  $-(1/4 +k^2)< -1/4$; this corresponds to the ``strongly attractive'' case which requires a boundary condition at $v=0$ to make $\hat{\H}_s$ self-adjoint. Here the boundary condition is
\begin{equation}
\int \dd \varphi \left[ v \bar\Psi\pdv{}{v}\Phi- v \Phi\pdv{}{v}\bar\Psi \right]_{v=0}^{v=\infty}=0
\label{bound-2}
\end{equation}
which is nontrivial only at $v=0$, contrary to what happened in \eqref{bound} where only the limit $v=\infty$ played a rôle in the form of the wavefunctions. For fixed $k$ one finds again a one-parameter family of possible subspaces of solutions satisfying the boundary condition, which then leads to a choice of free function once $k$ is treated as a dynamical variable. Overall  the normalisable wavefunctions are
\begin{eqnarray}
\fl
\Psi(v,\varphi,t)&=&\int_{-\infty}^{\infty} \frac{\dd k}{2\pi}e^{\im k\varphi}\left[ \sum_{n=-\infty}^\infty e^{\im  \frac{\lambda_n^k}{\hbar}t} B(k,\lambda_n^k) \frac{1}{\hbar}\sqrt{\frac{-2\lambda_n^k\sinh(k\pi)}{k\pi}}\K{\im k}{-\lambda_n^k} \right. \nonumber \\
\fl
&& +\left. \int_0^\infty \frac{\dd \lambda}{2\pi\hbar} e^{\im \frac{\lambda}{\hbar}t}A(k,\lambda) \frac{\sqrt{2\pi}\Re\left[ e^{\im \vartheta(k)-\im k \log\sqrt{\frac{\lambda}{\lambda_0}}}\J{\im k}{\lambda}\right]}{\sqrt{\hbar \cos\left(-2\vartheta(k)+k\log\frac{\lambda}{\lambda_0}\right)+\hbar \cosh(k\pi)}}\right]\, ,
\label{general-solution-2}
\end{eqnarray}
where $\vartheta(k)$ is the free function, $\lambda_0$ is an arbitrary reference scale and 
\begin{equation}
\lambda^k_n =-\lambda_0e^{-\frac{(2n+1)\pi}{k}+\frac{2\vartheta(k)}{k}}\,.
\end{equation}

We observe some similarities between the $\varphi$ clock theory constructed in this paper and this Schr\"odinger-like $t$ clock theory. For example, only real combinations of imaginary order Bessel functions are allowed and only certain discrete values of some parameters (here only certain negative $\lambda$ values for each $k$, in the $\varphi$ time theory only some $\kappa$ values for each $\lambda$) are allowed. However, there  are no real order Bessel function states in the Schrödinger theory, as these states are not normalisable, and the general solutions \eqref{general-solution} and \eqref{general-solution-2} look quite different. One of the most important differences between the theories is that when we choose $\varphi$ as a clock the boundary condition \eqref{bound} is relevant at $v=\infty$, but when $t$ is the clock the boundary condition \eqref{bound-2} is relevant only at $v=0$. Hence in the first case the different modes are reflected from infinity rather than from the singularity ($v=0$) as we found for the $t$ time theory in \cite{Gielen2020} (and Gryb and Th\'ebault had analysed in detail in \cite{Gryb2019,Gryb2019b,*Gryb2019c}). It is then already clear that the choice of clock is important for the physical interpretation of these theories, in particular regarding singularity resolution: choosing $t$ as a clock leads to generic singularity resolution, but we would not expect this in the case where $\varphi$ is the clock. On the other hand, when $\varphi$ is used as time states should not be able to reach infinite volume but instead ``bounce'' at some finite maximal volume. These expectations will be confirmed in our detailed numerical analysis below.

If $\log(v/v_0)$ is chosen as a relational clock the resulting quantum theory is very different from the other quantum theories. Here one again starts by writing the Wheeler--DeWitt equation in the form \eqref{wdw2} and notices that this is a second order differential equation in $\log(v/v_0)$. This variable can then also be used as a relational time; the appropriate inner product is of Klein--Gordon form
\begin{equation}
\fl
\braket{\Psi}{\Phi}_{v}=\im \int_{-\infty}^\infty \dd t \int_{-\infty}^\infty \dd \varphi \left[ \bar{\Psi}(v,\varphi,t) v\pdv{}{v}\Phi(v,\varphi,t)-\Phi(v,\varphi,t)v\pdv{}{v}\bar{\Psi}(v,\varphi,t)\right]\, .
\end{equation}
Conservation of this inner product in $v$ can be shown to be equivalent to self-adjointness of the operator
$\hat{\mathcal{O}}(v)=\left(-\hbar^2\frac{\partial^2}{\partial\varphi^2}-\im \hbar v^2\pdv{}{t} \right)$ for an $L^2$ inner product in $t$ and $\varphi$, but this operator is already self-adjoint, so all states of the form \eqref{generalwf} have a time-independent norm (states with real order Bessel functions are not normalisable in this theory either).

For a standard Klein--Gordon equation it is possible to take a ``square root'' on both sides to obtain two possible Schr\"odinger equations corresponding to the two square roots. Positive- or negative-frequency solutions of the former are then the solutions to one or the other of these Schr\"odinger equations. For this theory, however, there is no such interpretation as the operator $\hat{\mathcal{O}}(v)$ is time-dependent. Let us suppose that we can find an operator $\hat{\H}$ such that the Wheeler--DeWitt equation is implied by a Schrödinger equation in the time coordinate $u=\log(v/v_0)$,
\begin{equation}
\fl
-\hbar^2 \frac{\partial^2}{\partial u^2}\Psi(u,\varphi,t)=\hat{\O}(u)\Psi(u,\varphi,t)\quad\stackrel{?}{\Longleftarrow}\quad \im \hbar \pdv{}{u}\Psi(u,\varphi,t)=\hat{\H}(u) \Psi(u,\varphi,t)\, .
\end{equation}  
This would require that $\hat{\H}^2+\im\hbar\frac{\partial\hat{\H}}{\partial u}=\hat{\O}$ and hence, in a state where all expectation values are well-defined, that
\begin{equation}
\expval{\hat{\H}^2}+\im \hbar \pdv{}{u}\expval{\hat{\H}}=\expval{\hat{\O}}\, .
\label{semiclass-approx}
\end{equation}
Now assuming that all operators in question are self-adjoint, one reaches a contradiction because the right-hand side is real whereas the left-hand side is complex unless $\hat{\H}$ is time-independent, implying  that $\hat{\O}$ is also time-independent which is not the case.

Hence, this theory cannot be interpreted in terms of Schr\"odinger equations, which makes its semiclassical interpretation difficult. There is an approximate effective Schr\"odinger description valid semiclassically, which however requires a complex Schrödinger time. We explored this option in \cite{Gielen2020} following the ideas of \cite{Bojowald2011}. In summary, one needs to relax the assumption that $u$ is the clock of the associated Schrödinger theory. Consider instead a clock $\tau$ which only satisfies $\pdv{}{\tau}=\pdv{}{u}$. Writing $\tau=u+\delta$ and expanding \eqref{semiclass-approx} to first order in $\delta$ (assuming covariances are small) we find
\begin{equation}
\tau=u-\hbar \frac{\im }{2\expval{\hat{\H}}}
\end{equation}
which is complex. The imaginary part is relevant near the classical singularity but falls off at large $v$ (or $u$) as one would have expected. Hence, even though as a quantum theory this theory is much simpler than the other ones we have studied -- it does not require any boundary condition -- it is difficult to associate a notion of time evolution generated by a Hamiltonian to this theory, even semiclassically.

We can see how the choice of clock influences the mathematical and physical properties  of the resulting theory. Classically, none of the parameters $t$, $\varphi$ and $v$ are preferred clocks (as long as they are monotonic in the interval considered). However, the three quantum theories we found are formally very different. The main differences between the three theories are:
\begin{itemize}
\item \emph{Boundary conditions.} Demanding unitarity of the quantum theory when using either $\varphi$ or $t$ as a clock leads to the appearance of a nontrivial boundary condition \eqref{bound} or \eqref{bound-2}. Even though these boundary conditions look very similar, they do not have the same effect on the allowed wavefunctions as \eqref{bound} is only relevant at the classical singularity $v=0$ and \eqref{bound-2} is only relevant at $v=\infty$. These conditions ensure that the wavefunctions are reflected from these points and they imply that all allowed solutions can be written in terms of real combinations of Bessel functions. In both cases there is not a unique subspace of solutions to the boundary condition, instead the different subspaces one can choose are parametrised by free functions $\theta(k)$ and $\vartheta(\lambda)$. However, when $\log(v/v_0)$ is chosen as a clock the allowed states are only restricted by normalisability. The three theories are hence all very different in requiring a boundary condition only at small $v$, only at large $v$, or not at all. As a result they will all make very different physical predictions as to what kind of quantum behaviour is expected or generic.
\item \emph{Hamiltonian interpretation of the theory.} Also here, the three theories considered are all rather different. The quantum theory with $t$ clock is already in Schr\"odinger form and evidently formulated in terms of Hamiltonian evolution. In the theory discussed in the main part of the paper, we write the Wheeler--DeWitt equation as 
\begin{equation}
-\hbar^2\frac{\partial^2}{\partial\varphi^2}\Psi(v,\varphi,t)=\hat{\O}\Psi(v,\varphi,t)
\end{equation}
where $\hat{\O}$ is time-independent in the sense that it does not depend on $\varphi$. However, $\hat{\O}$ is not positive and so if one wanted to take a ``square root'' of this equation one would first need to exclude by hand the real exponential states \eqref{kappas}. In the third case where we use the volume $v$ as time, a Hamiltonian interpretation only exists semiclassically and requires the introduction of a complex time different from $v$.
\end{itemize}
Hence, we see that general covariance is clearly broken in the quantum theory: the choice of clock influences even the basic properties of the quantum theory in an essential way. We will see in more detail in the next section how this choice also influences the physical properties of the theory. It is worth mentioning that in loop quantum cosmology, models such as FLRW universes \cite{Pawlowski2011,Ashtekar2006,*Ashtekar2006b,*Ashtekar2006c}, Bianchi I \cite{Chiou2007} or Bianchi IX \cite{Wilson-Ewing2010} have all been quantised using a scalar field clock. These models show resolution of the big bang singularity, unlike the analogous quantum cosmology based on the Wheeler--DeWitt theory, nevertheless our analysis here suggests that a breaking of general covariance due to the choice of relational clock could also arise in loop quantum cosmology. It is not yet known whether singularity resolution is a generic feature of loop quantum cosmology or only arises due to the chosen relational clock, as also mentioned e.g.~in \cite{Bojowald2020}.

\section{Numerical results}
\label{numres}

One of the principal reasons to consider quantum gravity and quantum cosmology models is the hope to resolve the singularities that appear in general relativity, in particular the cosmological big bang singularity. However, to do this one first needs to specify the criteria one applies in order to claim singularity resolution. For discussion of some possible criteria see e.g.~\cite{Ashtekar2008,*Husain2004,*Kiefer2010}. We use a criterion similar to that proposed for the same model by Gryb and Th\'ebault \cite{Gryb2019,Gryb2019b,*Gryb2019c}, demanding that the expectation values of classically singular quantities (in particular, the volume or scale factor) are always non-singular. A different stronger criterion would be to demand that physical quantities such as the energy density have a universal upper bound satisfied for all states, as is the case in many models of loop quantum cosmology, but we will be satisfied with non-singular behaviour for any given state (in which for example the maximum energy density can be state-dependent). In this section we will define semiclassical states for the theory constructed in sections \ref{phi-clock} to \ref{phi-summary}, show plots of the relevant expectation values and finally compare our new results (with $\varphi$ as a clock) to previous results of \cite{Gielen2020} where the clock variable was either $t$ or $v$. The plots are obtained using the software \texttt{Mathematica}.

In section \ref{classth} we have given explicit expressions for the classical Dirac observables $t(\varphi=\varphi_1)$ and $v(\varphi=\varphi_1)$, corresponding to the value of $t$ or $v$ when the scalar field $\varphi$ takes a given value $\varphi_1$. We can compare these classical Dirac observables to the quantum expectation values $\expval{t(\varphi)}_{\Psi_{sc}}$ and $\expval{v(\varphi)}_{\Psi_{sc}}$ for a given semiclassical state $\Psi_{sc}$. Since the classical singularity is at $v=0$, if we find that $\expval{v(\varphi)}_{\Psi_{sc}} >C_{\Psi_{sc}}>0$ for some $C_{\Psi_{sc}}$ (which can be state-dependent) for all $\varphi$, then the evolution will be considered nonsingular. We interpret any failure to observe such behaviour as singular. 

The general form of a normalised wavefunction was given in \eqref{general-solution}. Here for concreteness we restrict ourselves to the $\lambda>0$ sector and we do not include any of the zero-norm modes characterised by the functions $\gamma^+_n(\lambda)$ and $\gamma^-_n(\lambda)$. We also set the free function $\kappa_0(\lambda)$ characterising self-adjoint extensions to zero for simplicity. Hence the only non-zero free function left in \eqref{general-solution} is $\alpha(k,\lambda)$. We choose $\lambda>0$ since in any perfect fluid interpretation $\lambda$ is proportional to the energy density of the fluid, hence $\lambda>0$ ensures that we are dealing with non-exotic matter. This includes the case of dark energy, for which $\lambda>0$ would match our own Universe. Perhaps most importantly, this case is the most interesting one because of the presence of the nontrivial boundary condition \eqref{bound}. Furthermore, again for simplicity, we only include modes with $k>0$. While $\kappa_0(\lambda)\equiv 0$ is the simplest choice of self-adjoint extension, one may ask whether the choice of self-adjoint extension changes the results significantly. In \cite{Pawlowski2011} Paw{\l}owski and Ashtekar studied our model at fixed $\lambda>0$ and found that the quantum evolution was not sensitive to the choice of self-adjoint extension; we expect similar behaviour here but have not explicitly verified this for all possible choices.

 Our semiclassical state is defined by
\begin{equation}
\fl
\Psi_{sc}(v,\varphi,t)=\int_{0}^{\infty} \dk{k} \int_{0}^{\infty}\dl{\lambda} e^{\im k\varphi}e^{\im \frac{\lambda}{\hbar}t} \alpha_{sc}(k,\lambda) \sqrt{\frac{2\pi}{\sinh(\abs{k}\pi)}}\Re\left[\J{\im |k|}{\lambda}  \right]\, ,
\label{semiclass}
\end{equation} 
where for $\alpha_{sc}(k,\lambda)$ we take a normalised Gaussian in $k$ and $\lambda$ centred around classical parameters $k_c$ and $\lambda_c$,
\begin{equation}
\alpha_{sc}(k,\lambda)=C\frac{2\sqrt{\hbar\pi}}{\sqrt{\sigma_k\sigma_\lambda}}e^{-\frac{(k-k_c)^2}{2\sigma_k^2}}e^{-\frac{(\lambda-\lambda_c)^2}{2\sigma_\lambda^2}}\, .
\label{alpha-sc}
\end{equation}
 The parameters $\sigma_\lambda$ and $\sigma_k$ are the standard deviation of the Gaussians. The constant $C>0$ ensures that we have $\int_0^\infty \dl{\lambda}\int_{0}^{\infty} \dk{k} \abs{\alpha_{sc}(k,\lambda)}^2=1$, but if $k_c$ and $\lambda_c$ are more than a few standard deviations away from zero we have $C\approx 1$ to very good approximation.

\subsection{Numerical analysis of expectation values $\expval{v(\varphi)}_{\Psi_{sc}}$ and $\expval{t(\varphi)}_{\Psi_{sc}}$}
\label{num-phi}

We start by analysing $\expval{v(\varphi)}_{\Psi_{sc}}$. The expression for this expectation value is
\begin{eqnarray}
\fl
\expval{v(\varphi)}_{\Psi_{sc}}&=&\int_0^\infty \frac{\dd k_1 \dd k_2}{(2\pi)^2}\int_0^\infty \dl{\lambda} \int_0^\infty\dd v  \frac{2\pi e^{\im(k_2-k_1)\varphi}(k_1+k_2)}{\sqrt{\sinh(\abs{k_1}\pi)\sinh(\abs{k_2}\pi)}} \times \nonumber\\
\fl
&& \bar{\alpha}_{sc}(k_1,\lambda)\alpha_{sc}(k_2,\lambda)\Re\left[\J{\im \abs{k_1}}{\lambda} \right]\Re\left[\J{\im \abs{k_2}}{\lambda} \right]\, .
\end{eqnarray}
Recall that we are using the redefinition of the inner product presented in \eqref{inner-prod-mod}, which ensures the state has positive norm. We maintain the notation $\abs{k}$ rather than $k$ so that the result could be straightforwardly extended to $k<0$. We can now calculate the $v$ integral analytically, which greatly speeds up the numerics. The relevant formula to use is \eqref{digamma_integral}. The integral diverges logarithmically, hence, in order to obtain a finite result, we introduce a finite cutoff $\Xi_v$. With this cutoff the expectation value is
\begin{eqnarray}
\fl
\expval{v(\varphi)}_{\Psi_{sc}}&\approx &-\int \frac{\dd k_1 \dd k_2}{(2\pi)^2} \frac{\dd \lambda}{2\pi \hbar} \frac{\hbar}{\sqrt{\lambda}}\frac{e^{\im(k_2-k_1)\varphi}(k_1+k_2)\cosh(\frac{k_1\pi}{2})\cosh(\frac{k_2\pi}{2})}{\sqrt{\sinh(\abs{k_1}\pi)\sinh(\abs{k_2}\pi)}}\bar{\alpha}_{sc}(k_1,\lambda)\alpha_{sc}(k_2,\lambda) \times \nonumber \\
\fl
&&\left\lbrace \log\left(\frac{4\hbar^2}{\lambda \Xi_v^2}\right)+\psi\left(\frac{1}{2}(1-\im(\abs{k_1}-\abs{k_2}) )\right)+\psi\left(\frac{1}{2}(1+\im(\abs{k_1}-\abs{k_2}))\right)\right. \nonumber \\
\fl
&& \left. +\psi\left(\frac{1}{2}\left(1+\im (\abs{k_1}+\abs{k_2})\right)\right)+\psi\left(\frac{1}{2}\left(1-\im(\abs{k_1}+\abs{k_2})\right)\right)+2\gamma\right\rbrace\, ,
\label{expval-v}
\end{eqnarray}
where $\gamma$ refers to the Euler--Mascheroni constant and $\psi(x)$ to the digamma function. The result explicitly depends on the value of the cutoff $\Xi_v$ but as this dependence is logarithmic, drastic changes of the cutoff only lead to mild changes in the results. Below we compare different values of the cutoff, and take it to values as high as $10^{10}$. There is a quantitative difference to the results especially around $\varphi=0$ but it is clear that qualitative features of the expectation value $\expval{v(\varphi)}_{\Psi_{sc}}$ (which we are mostly interested in here) are not too sensitive to it. Numerical evaluation of the $v$ integral (without using \eqref{digamma_integral}) would require a similar cutoff at large $v$, leading to a similar ambiguity.

The expectation value \eqref{expval-v} is symmetric under $\varphi\rightarrow -\varphi$, hence we compare it to the classical solution
\begin{equation}
v_c(\varphi)=\frac{\hbar k_c}{\sqrt{\lambda_c}\abs{\sinh(\varphi)}}\, ,
\end{equation}
which corresponds to \eqref{fcts-of-phi} where $\varphi_0=0$, and with $\pi_\varphi=\hbar k_c$. 

The integral \eqref{expval-v} is still very hard to compute numerically, and a further approximation is necessary in order to obtain fast and reliable results. Notice that the necessary integrals over $\lambda$ are relatively simple, corresponding to 
\begin{equation}
\frac{2\hbar\sqrt{\pi}}{\sigma_\lambda} \int\frac{\dd \lambda}{2\pi \hbar} \frac{\hbar}{\sqrt{\lambda}} e^{-\frac{(\lambda-\lambda_c)^2}{\sigma_\lambda^2}}\,,\quad \frac{2\hbar\sqrt{\pi}}{\sigma_\lambda}  \int\frac{\dd \lambda}{2\pi \hbar} \frac{\hbar}{\sqrt{\lambda}} e^{-\frac{(\lambda-\lambda_c)^2}{\sigma_\lambda^2}}\log\left(\frac{4\hbar^2}{\lambda \Xi_v^2}\right)\,,
\end{equation}
which can again be evaluated analytically, but the resulting expression is very complicated and will not be given here. If we consider the limit in which $\sigma_\lambda$ is extremely small, we can check that these integrals reduce, as one would expect, to the value of the integrand evaluated at $\lambda=\lambda_c$. In the following we will use this approximation of very small $\sigma_\lambda$ and replace all $\lambda$ integrals by the integrand evaluated at $\lambda=\lambda_c$. Since the inner product used here means that different $\lambda$ sectors are decoupled, it makes sense to reduce to this limit of effectively only a single $\lambda$. This simple approximation speeds up the numerics and  reduces integration errors. 

\begin{figure}[h!]
\centering 
\includegraphics[scale=0.5]{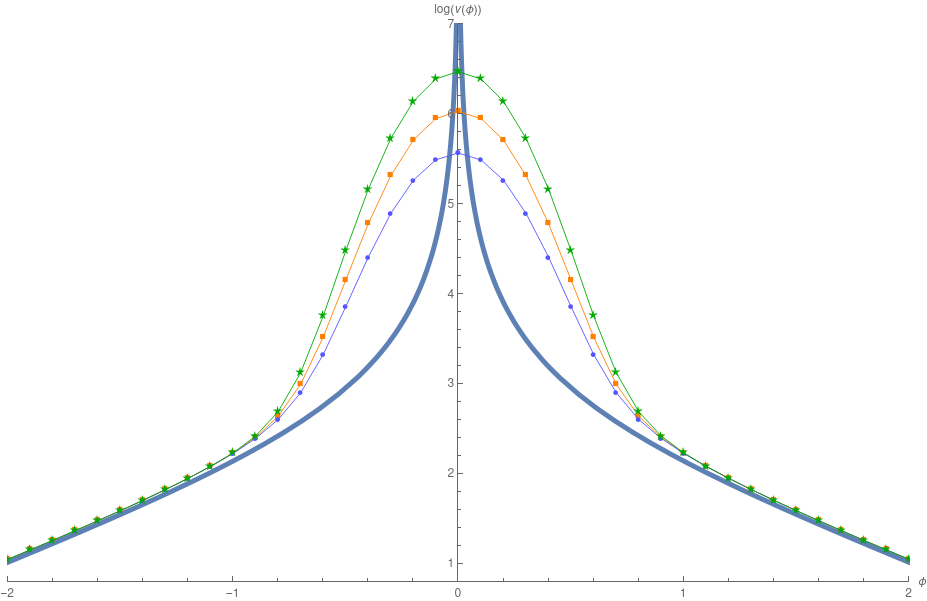}
\caption{Quantum solution $\log{\expval{v(\varphi)}_{\Psi_{sc}}}$ (discrete shapes) and classical solution $\log{v_c(\varphi)}$ (solid line) for the parameters $k_c=10$, $\sigma_k=3$, $\lambda_c=1$ and $\Xi_v=10^5$ (blue dots), $\Xi_v=10^7$ (orange squares), $\Xi_v=10^{10}$ (green stars). The points have been linked to facilitate the reading of the figure. We work in units $\hbar=1$.}
\label{v-phi}
\end{figure}
Numerical results for $\expval{v(\varphi)}_{\Psi_{sc}}$ for different values of the cutoff are presented in figure \ref{v-phi}. We observe that for values of $\abs{\varphi}\gtrsim 1$, both the classical curve and the quantum expectation value agree very closely. At smaller values of $\abs{\varphi}$, the two curves start diverging; the quantum expectation value reaches a finite maximum  at $\varphi=0$ and smoothly transitions between the expanding and the collapsing branch of the classical solution. This behaviour is as we would have expected from the analysis of section \ref{qtumth}, since we had to impose a reflective boundary condition at $v=\infty$, which corresponds to $\varphi=0$. However, the singularity $v=0$ is still present: figure \ref{v-phi} shows that at large $\abs{\varphi}$ the expectation value becomes smaller and smaller, following exactly the classical solution. We also see that a large cutoff increases the values of $\expval{v(\varphi)}_{\Psi_{sc}}$ for small $\varphi$; however the numerical results are of the same order of magnitude,  exhibiting again the relatively weak logarithmic dependence on the cutoff. Independently of the cutoff we see that when quantum corrections first become important as the universe expands, they lead to a faster expansion than in the classical theory, only to then make the expansion slow down and stop at a finite volume. This first phase of more rapid expansion is in agreement with the results of \cite{Bojowald2010b}, where a systematic expansion into higher order quantum fluctuations around the classical trajectory was studied in a similar model (again, with a fixed $\lambda>0$). In this approach one sees explicitly how quantum fluctuations diverge as the volume grows, which then trigger the recollapse we see here.

There is a more quantitative argument to show that the classical singularity is not resolved in this theory. If we consider the integral (\ref{expval-v}) for very large $|\varphi|$, we see that in this limit the integral becomes infinitely peaked around $k_1=k_2$. More concretely, we can argue that in a distributional sense
\begin{equation}
\lim_{\varphi\rightarrow \pm\infty}e^{\im(k_1-k_2)\varphi}=\pm \im \pi(k_1-k_2)\delta(k_1-k_2)\, ,
\label{phi-infty}
\end{equation}
which when substituted into (\ref{expval-v}) leads to the conclusion $\lim_{\varphi \rightarrow \pm \infty} \expval{v(\varphi)}_{\Psi_{sc}}=0$. (\ref{phi-infty}) follows from
\begin{equation}
\lim_{\varphi\rightarrow \pm\infty}\int \frac{{\rm d}k}{k} e^{\im k\varphi}f(k)=\lim_{\varphi\rightarrow \pm\infty}\pm\int \frac{{\rm d}\kappa}{\kappa} e^{\im \kappa}f\left(\frac{\kappa}{\varphi}\right)
\end{equation}
whose imaginary part yields $\pm\im\pi f(0)$ while the real part must be a distribution that only depends on the odd part of a test function $f$ but also only on its value at zero, so that it must be zero itself.

\begin{figure}[h!]
\centering 
\includegraphics[scale=0.48]{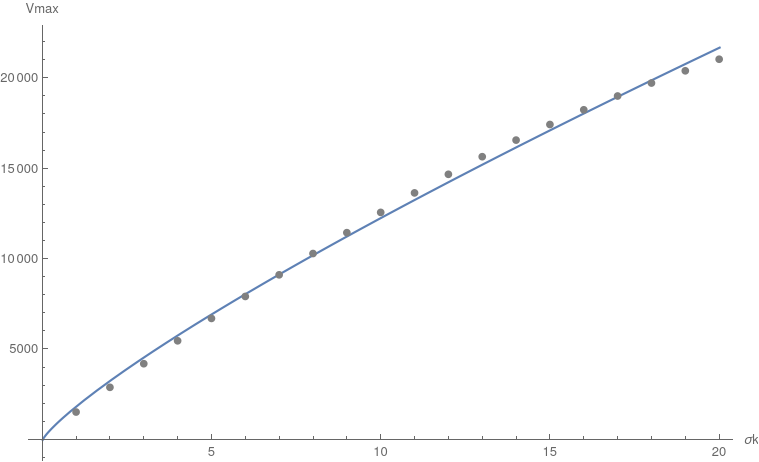}
\caption{Values of the maximum volume as function the standard deviation $\sigma_k$ for $k_c=100$, $\lambda=1$ and $\Xi_v=10^8$. The solid line is a fit of the form $a\sigma_k^b$ where $a\approx 1853$ and $b\approx 0.8212$. Here a larger $k_c$ was chosen to allow higher values for $\sigma_k$. }
\label{v-max}
\end{figure}

\begin{figure}[h!]
\centering 
\includegraphics[scale=0.48]{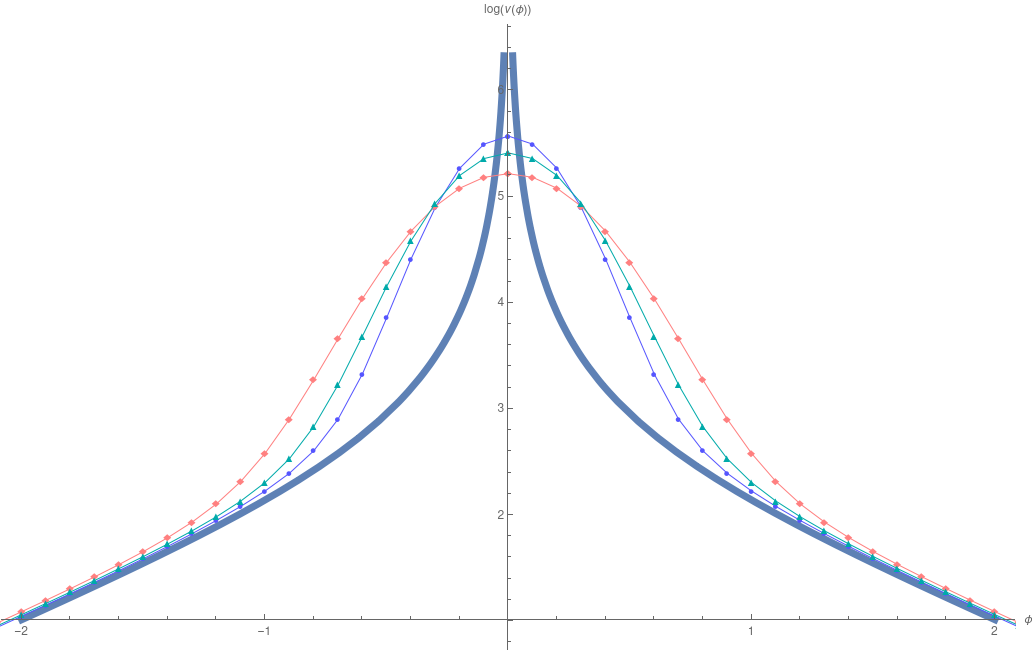}
\caption{Quantum solution $\log\expval{v(\varphi)}_{\Psi_{sc}}$ (discrete shapes) and classical solution $\log v_c(\varphi)$ (continuous blue line) for different values of the standard deviation $\sigma_k$. The blue circles correspond to $\sigma_k=3$, the cyan triangles to $\sigma_k=2.5$ and the pink diamonds to $\sigma_k=2$. The points have been joined by lines to facilitate the reading. The remaining parameters are $k_c=10$, $\lambda_c=1$ and $\Xi_v=10^5$ (again with $\hbar=1$).}
\label{v-sigmas}
\end{figure}

Another interesting feature of this model is that the maximum volume $\expval{v(0)}_{\Psi_{sc}}$ grows with the standard deviation $\sigma_k$, as shown in figure \ref{v-max}. Moreover, for larger $\sigma_k$ the classical and quantum solution agree up to higher volume and there is a more sudden transition between the two branches, as shown in figure \ref{v-sigmas}. This perhaps surprising behaviour can be explained by remembering $k$ and $\varphi$ are conjugate, so a wide spread in $k$ implies smaller uncertainty in $\varphi$, leading to the observed curves. Similar behaviour was already observed in \cite{Gielen2020, Gryb2019} for the conjugate pair $t$ and $\lambda$. 

The most interesting feature we observe here is a quantum recollapse at $\varphi=0$ where the universe reaches a maximum volume. In order to verify that this result accurately describes the quantum dynamics  and that the quantum transition between the expanding and the collapsing branch is continuous, we have calculated the difference $\expval{v(0)}_{\Psi_{sc}}-\expval{v(\epsilon)}_{\Psi_{sc}}$ for $\epsilon\rightarrow 0$. The results are presented in Table \ref{table}. We see that as $\varphi$ tends to 0, the expectation value converges very quickly (and presumably continuously) to its value at $\varphi=0$, confirming that $\expval{v(\varphi)}_{\Psi_{sc}}$ is well defined for all $\varphi$.
\begin{table}[h!]
\begin{center}
\begin{tabular}{c|c}
$\varphi$ & $\log[\expval{v(0)}_{\Psi_{sc}}]-\log[\expval{v(\varphi)}_{\Psi_{sc}}]$ \\
\hline
$10^{-1}$ & $8.77 \times 10^{-2}$\\
$10^{-3}$ & $8.76 \times 10^{-6}$ \\
$10^{-5}$ & $\ 8.76 \times  10^{-10}$
\end{tabular}
\caption{Difference between  $\expval{v(0)}_{\Psi_{sc}}$ and nearby values for $k_c=10$, $\sigma_k=3$, $\lambda_c=1$, and $\Xi_v=10^5$, with $\hbar=1$.}
\label{table}
\end{center}
\end{table}

To complete our analysis of the theory we also study $\expval{t(\varphi)}_{\Psi_{sc}}$. Given our state $\Psi_{sc}$ this expectation value is given by
\begin{eqnarray}
\fl
\expval{t(\varphi)}_{\Psi_{sc}}&=&\int_0^\infty \frac{\dd k_1 \dd k_2}{(2\pi)^2}\int_0^\infty \frac{\dd \lambda_1 \dd \lambda_2}{(2\pi \hbar)^2} \int_0^\infty \frac{\dd v}{v} \int_{-\infty}^{\infty} \dd t \, \frac{2\pi e^{\im(k_2-k_1)\varphi}(k_1+k_2)}{\sqrt{\sinh(\abs{k_1}\pi)\sinh(\abs{k_2}\pi)}} \times \nonumber\\
\fl
&&t e^{\im(\lambda_2-\lambda_1)\frac{t}{\hbar}} \bar{\alpha}_{sc}(k_1,\lambda_1)\alpha_{sc}(k_2,\lambda_2)\Re\left[\J{\im \abs{k_1}}{\lambda_1} \right]\Re\left[\J{\im \abs{k_2}}{\lambda_2} \right]\, .
\end{eqnarray}
This expectation value is not as straightforward to calculate because of the extra factor of $t$. However, we can use the fact that
\begin{eqnarray}
\fl
\int \frac{\dd \lambda_1 \dd \lambda_2}{(2\pi \hbar)^2}\dd t  \ t\, e^{\im(\lambda_2-\lambda_1)\frac{t}{\hbar}}F(\lambda_1, \lambda_2)=-\frac{\im \hbar}{2}\int \dl{\lambda} \left[\pdv{}{\lambda_1}F(\lambda_1,\lambda_2)-\pdv{}{\lambda_2}F(\lambda_1,\lambda_2) \right]_{\lambda_1=\lambda_2=\lambda}
\end{eqnarray}
to write this expectation value as  a combination of two contributions,
\begin{eqnarray}
\fl
\expval{t(\varphi)}_{\Psi_{sc}}&=&-\im \pi\hbar\int \frac{\dd k_1 \dd k_2}{(2\pi)^2}\dl{\lambda} \frac{\dd v}{v}\frac{ e^{\im(k_2-k_1)\varphi}(k_1+k_2)}{\sqrt{\sinh(\abs{k_1}\pi)\sinh(\abs{k_2}\pi)}} \times \nonumber \\
\fl
&&\left\lbrace \left(\alpha_{sc}(k_2,\lambda)\partial_\lambda\bar{\alpha}_{sc}(k_1,\lambda)-\bar{\alpha}_{sc}(k_1,\lambda)\partial_\lambda\alpha_{sc}(k_2,\lambda) \right)\Re\left[\J{\im \abs{k_1}}{\lambda} \right]\Re\left[\J{\im \abs{k_2}}{\lambda} \right]\right.\nonumber \\
\fl 
&& + \bar{\alpha}_{sc}(k_1,\lambda)\alpha_{sc}(k_2,\lambda)H(k_1,k_2,\lambda,v)  \Bigg\rbrace
\label{t-phi-1}
\end{eqnarray}
with
\begin{eqnarray}
\fl
H(k_1,k_2,\lambda,v)&=&\frac{v}{ 4\hbar \sqrt{\lambda}}\left\lbrace \Re\left[\J{\im \abs{k_1}}{\lambda} \right]\Re\left[\J{1+\im \abs{k_2}}{\lambda}-\J{-1+\im \abs{k_2}}{\lambda}\right]\right. \nonumber \\
\fl
&&+ \left.  \Re\left[\J{\im \abs{k_2}}{\lambda}\right]  \Re\left[\J{-1+\im \abs{k_1}}{\lambda}-\J{1+\im \abs{k_1}}{\lambda}\right] \right\rbrace\, .
\end{eqnarray}
The second line in \eqref{t-phi-1} vanishes if $\alpha_{sc}(k,\lambda)$ is real and separable in $k$ and $\lambda$ which is our case. Hence, only the last line is important. Once again we can first integrate over $v$; $H(k_1,k_2,\lambda,v)$ is a sum of 16 terms which lead to 16 integrals of the form \eqref{H-int-1} or \eqref{H-int-2}. After doing all the integrals and summing over the 16 terms one finds that the terms multiplying $\delta(\abs{k_1}\pm\abs{k_2})$ all cancel, hence only the contribution at $v=\infty$ is important. There some terms diverge as $\log(\frac{4\hbar^2 }{\lambda v^2})$, but these terms also cancel. The terms containing digamma functions also simplify and one obtains 
\begin{eqnarray}
\int \frac{\dd v}{v} H(k_1,k_2,\lambda,v)&=& \frac{1}{4\lambda}\left(\coth\left((\abs{k_1}+\abs{k_2})\frac{\pi}{2}\right)\sinh\left((\abs{k_1}-\abs{k_2})\frac{\pi}{2}\right)\right.\nonumber\\
&&+\left. \coth\left((\abs{k_1}-\abs{k_2})\frac{\pi}{2}\right)\sinh\left((\abs{k_1}+\abs{k_2})\frac{\pi}{2}\right) \right)\, .
\end{eqnarray}

Hence, the final result for the expectation value $\expval{t(\varphi)}_{\Psi_{sc}}$ is
\begin{eqnarray}
\fl
\expval{t(\varphi)}_{\Psi_{sc}}&=&-\frac{\im \pi\hbar}{4}\int \frac{\dd k_1 \dd k_2}{(2\pi)^2} \dl{\lambda} \frac{ e^{\im(k_2-k_1)\varphi}(k_1+k_2)}{\lambda\sqrt{\sinh(\abs{k_1}\pi)\sinh(\abs{k_2}\pi)}} \bar{\alpha}_{sc}(k_1,\lambda)\alpha_{sc}(k_2,\lambda)\times 
 \label{expval-t}
\\
\fl
&&\left(\coth\left(\frac{k_+\pi}{2}\right)\sinh\left(\frac{k_-\pi}{2}\right)+\coth\left(\frac{k_-\pi}{2}\right)\sinh\left(\frac{k_+\pi}{2}\right) \right)\,,\; k_\pm:=\abs{k_1}\pm\abs{k_2}\,.  \nonumber
\end{eqnarray}
This expression is antisymmetric with respect to the change $\varphi\rightarrow -\varphi$ which motivates comparing this quantum expectation value to the classical solution
\begin{equation}
t_c(\varphi)=-\frac{\hbar k_c}{2\lambda_c}\coth(\varphi)\,
\label{t-classical}
\end{equation}
which corresponds to \eqref{fcts-of-phi} where $t_0=\varphi_0=0$. We stress that \eqref{expval-t} does not depend on any cutoff, unlike \eqref{expval-v}, so we can now proceed to evaluate \eqref{expval-t} numerically. As in the previous case of the integral \eqref{expval-v}, we now simplify matters by assuming the variance $\sigma_\lambda$ to be extremely small. Here the $\lambda$ dependent parts under the integral in \eqref{expval-t} are proportional to $1/\lambda$ times a Gaussian, and we cannot perform the integral analytically. But just as in the previous case we will implement the limit $\sigma_\lambda \rightarrow 0$ by replacing the $\lambda$ integral by the integrand evaluated for $\lambda=\lambda_c$. In this limit, the numerical analysis can be performed and gives meaningful results.

\begin{figure}[h!]
\centering
\includegraphics[scale=0.5]{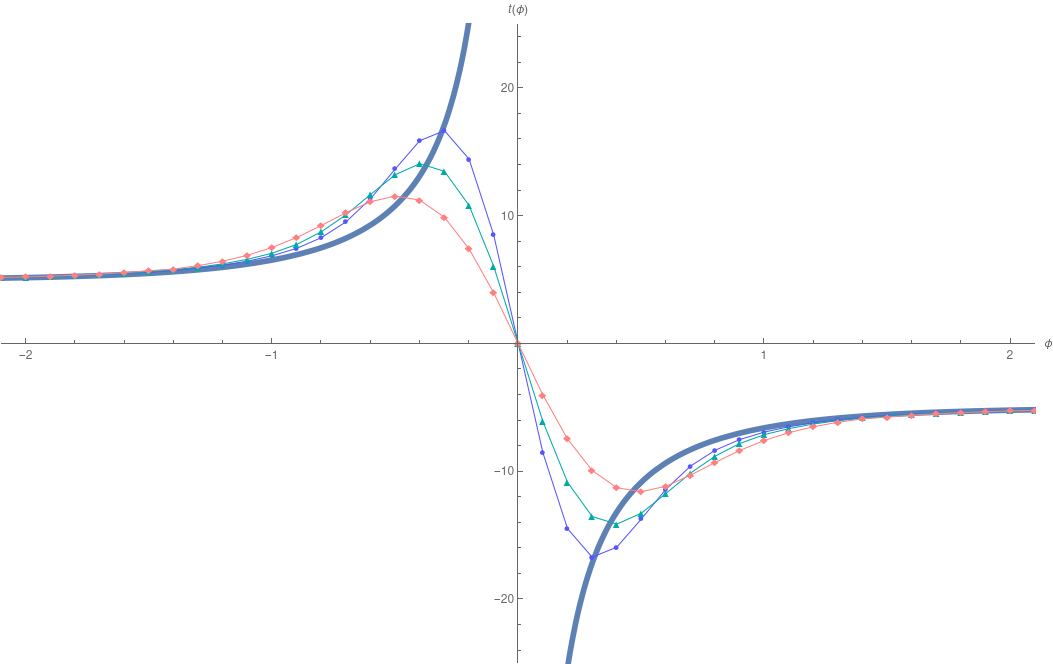}
\caption{Expectation values $\expval{t(\varphi)}_{\Psi_{sc}}$ (discrete shapes) and classical solution $t_c(\varphi)$ (continuous blue line) for different values of the standard deviation $\sigma_k$. The blue circles correspond to $\sigma_k=3$, the cyan triangles to $\sigma=2.5$ and the pink diamonds to $\sigma_k=2$. The remaining parameters take values $k_c=10$ and $\lambda_c=1$. Again we take $\hbar=1$.}
\label{t-sigmas}
\end{figure}

In figure \ref{t-sigmas} we present the numerical results for the expectation value $\expval{t(\varphi)}_{\Psi_{sc}}$ in comparison to the classical solution (\ref{t-classical}). Once again, we observe that for $\abs{\varphi}> 1$ the two curves agree very closely. For small $\abs{\varphi}$ the quantum expectation value reaches an extremum and then goes to 0, to transition smoothly between the classical expanding and contracting branch. We also observe the same behaviour with respect to changes in the standard deviation of the Gaussian: for bigger $\sigma_k$, the classical and quantum solution agree more closely and the transition between the expanding and contracting branch is more abrupt. These figures also show that the expectation value $\expval{t(\varphi)}_{\Psi_{sc}}$ is no longer monotonic with respect to $\varphi$ in any of the two sectors $t<0$ and $t>0$ and experiences a turnaround, unlike what happens classically.

We can also give an analytical expression for the quantum expectation value in the limit of large $\abs{\varphi}$ using \eqref{phi-infty}; we find
\begin{eqnarray}
\lim_{\varphi \rightarrow \pm \infty}\expval{t(\varphi)}_{\Psi_{sc}}= \mp \frac{\hbar}{2}\int \dk{k}\dl{\lambda}\frac{k}{\lambda}\abs{\alpha_{sc}(k,\lambda)}^2
\end{eqnarray}
Given that $\alpha_{sc}(k,\lambda)$ is a normalised Gaussian in $k$ and $\lambda$, the result of the $k$ integral is simply the mean, $k_c$. We also take the result for the $\lambda$ integral to be $1/\lambda_c$ (which requires regularising the logarithmic divergence at $\lambda=0$). Hence we find
\begin{equation}
\lim_{\varphi\rightarrow\pm \infty }\expval{t(\varphi)}_{\Psi_{sc}}= \mp \frac{\hbar k_c}{2\lambda_c}=\lim_{\varphi \rightarrow \pm \infty} t_c(\varphi)\, 
\end{equation}
and once again for large values of $\abs{\varphi}$ the quantum expectation value tends to the classical solution both analytically and numerically.  Since large $\abs{\varphi}$ corresponds to the classical big bang/big crunch singularity this again illustrates that there is no singularity resolution as the quantum expectation value follows the classical solution.

\begin{figure}[h!]
\centering
\includegraphics[scale=0.6]{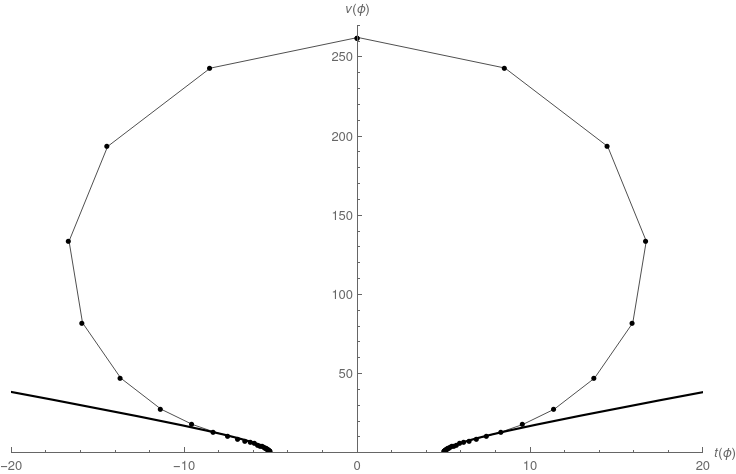}
\caption{Expectation values $\expval{t(\varphi)}_{\Psi_{sc}}$ with respect to $\expval{v(\varphi)}_{\Psi_{sc}}$ (dots) and classical solution $v(t)$ (solid line) for $k_c=10$, $\lambda_c=1$, $\sigma=3$ and $\Xi_v=10^5$, with $\hbar=1$.}
\label{v-vs-t}
\end{figure}

In figure \ref{v-vs-t} we give a parametric plot of $\expval{v(\varphi=\varphi_\im)}_{\Psi_{sc}}$ with respect to $\expval{t(\varphi=\varphi_\im)}_{\Psi_{sc}}$ in comparison to the classical curve $v(t)$ for $\varphi_\im\in[-3,3]$. The quantum universe emerges from the classical singularity and is very close to the classical curve evolving forward in $t$. Then $t$ has a turnaround, starting to go backwards and the universe reaches a maximum volume when $t=0$ to then approach the classical contracting solution while $t$ starts going forward again. Again it is clear that the classical singularity is not resolved. 

In conclusion, the $\varphi$ theory has very interesting dynamics: the universe follows the classical expanding solution, but instead of expanding up to infinity like in the classical theory it reaches a finite maximum value at $\varphi=0$ and then recollapses. The ``time coordinate'' $t$ also suffers from large quantum effects, transitioning  smoothly from the $t>0$ branch to the $t<0$ one and therefore losing its injectivity. Since $v\rightarrow 0$ when $\varphi\rightarrow \pm \infty$, the big bang and big crunch singularities are still present.

\subsection{Comparison with other choices of relational clock}

We have seen in section \ref{comparsec} that the classical covariance under time reparametrisations is broken in the quantum theory since different relational clocks require different boundary conditions. To see more explicitly the implications of this breaking of covariance, in this section we compare the results obtained in section \ref{num-phi} with our previous work \cite{Gielen2020}, in which either $t$ or $v$ (or equivalently, $\log(v/v_0)$) were used as clock. We present the main similarities and differences of the theories, and give conformal diagrams for a better visualisation of the resulting picture for a quantum spacetime history. 

Let us begin with the simplest theory in which $v$ is used as a clock. Here there are no nontrivial boundary conditions, and any solution to the Wheeler--DeWitt equation
\begin{eqnarray}
\fl
\Psi(v,\varphi,t)&=&\int_{-\infty}^\infty \dl{\lambda} \int_{-\infty}^\infty \dk{k} \sqrt{\frac{\pi}{2\sinh(\abs{k}\pi)}}e^{\im k\varphi}e^{\im \frac{\lambda}{\hbar}t}\times \nonumber \\
\fl
&&\left[ a(k,\lambda)\J{\im\abs{ k}}{\lambda} +b(k,\lambda)\J{-\im \abs{k}}{\lambda}\right]
\label{generalwave}
\end{eqnarray}
is allowed as long as $a(k,\lambda)$ and $b(k,\lambda)$ satisfy $\int \dl{\lambda} \dk{k} \left[\abs{a(k,\lambda)}^2+\abs{b(k,\lambda)}^2\right]=1$ for the state to be normalised. The asymptotic expression for small arguments of the Bessel functions of the first kind is
\begin{equation}
\J{\pm \im \abs{k}}{\lambda}\longrightarrow \frac{e^{\pm \im \abs{k}\log\left(\frac{\sqrt{\lambda}}{2\hbar}v\right) }}{\Gamma(1\pm \im \abs{k})}\,, \quad v\rightarrow 0\, .
\end{equation}
Hence, at small values of $v$, the general solution (\ref{generalwave}) is a combination of plane waves outgoing from the classical singularity and incoming to the singularity. We already saw that in the other cases where $t$ or $\varphi$ are the clock, the boundary condition only allows solutions built from real linear combinations of these two types of waves, but here one can choose the relative weight of outgoing and incoming modes freely. To underline this different behaviour of the $v$ theory, in \cite{Gielen2020} we focused on solutions that are purely outgoing, with $a(k,\lambda)=0$. As in the other cases, we also restricted ourselves to positive $k$ and $\lambda$, and chose for $b(k,\lambda)$ a Gaussian in $k$ and $\lambda$ with some standard deviation $\sigma_\lambda$ and (in this case) extremely small $\sigma_k$. The Gaussian was defined to be centred around some classical values $k_c$ and $\lambda_c$.

For this theory, we were able to obtain analytical expressions for the most interesting expectation value $\expval{t(v)}_{\Psi_{sc}}$. Namely, in a limit where the variance $\sigma_\lambda$ is also very small and at small $v$, we could approximate 
\begin{equation}
\expval{t(v)}_{\Psi_{sc}} \approx\frac{\hbar \abs{k_c}}{2\lambda_c} + \frac{v^2}{4 \hbar \abs{k_c}}-\frac{\lambda_c v^4}{16\hbar^3 (\abs{k_c}+\abs{k_c}^3)} + O(v^6)\, .
\end{equation}
This expectation value can be compared with the classical solution
\begin{equation}
t_c(v)=\frac{\hbar \abs{k_c}}{2\lambda_c} + \frac{v^2}{4 \hbar \abs{k_c}}-\frac{\lambda_c v^4}{16\hbar^3 \abs{k_c}^3} + O(v^6)
\label{t-v-class}
\end{equation}
and so near the classical singularity $v=0$ classical and quantum solutions only start diverging at $O(v^4)$.

\begin{figure}[!h]
\centering
\includegraphics[scale=0.42]{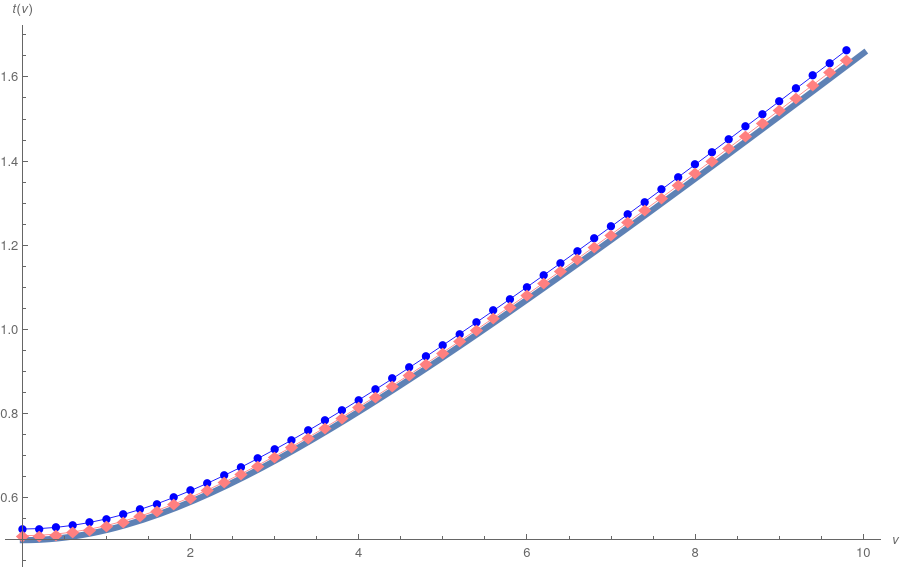}
\caption{Expectation value $\expval{t(v)}_{\Psi_{sc}}$  compared to classical solution $t_c(v)$ (solid line) for $k_c=10$, $\lambda=10_c$, with $\hbar=1$. The blue circles correspond to the quantum solution for $\sigma_\lambda=3$ and the pink diamonds correspond to the quantum solution for $\sigma_\lambda=2$.}
\label{t-v-plot}
\end{figure}
Full results for $\expval{t(v)}_{\Psi_{sc}}$ for a representative choice of parameters are shown in figure \ref{t-v-plot}. We see that the classical solution and the quantum expectation values always remain very close, which is of course in sharp contrast to the behaviour of the theory studied in section \ref{num-phi} where $\varphi$ was the clock. Neither of these two theories resolve the classical singularity, but if $v$ is chosen as clock there are quantum states for which corrections to the classical solution are very small throughout the whole evolution. Another observation is that figure \ref{t-v-plot} shows better agreement between quantum and classical solution for smaller values of $\sigma_\lambda$, whereas in the other theories we generally observe closer agreement for Gaussians with large variance.

Let us now focus on the ``Schr\"odinger'' theory for which $t$ is the clock. We gave the general normalisable solution in this theory in \eqref{general-solution-2} but in \cite{Gielen2020} again we focused on states of positive $k$ and $\lambda$. We also chose the free function appearing in the  self-adjoint extension to be $\vartheta(k)=0$. Our semiclassical states were hence of the form
\begin{eqnarray}
\fl
\Psi_{sc}(v,\varphi,t)=\int_0^\infty \dl{\lambda}\int_0^\infty \dk{k}e^{\im k\varphi}e^{\im \frac{\lambda}{\hbar}t} A_{sc}(k,\lambda) \frac{\sqrt{2\pi}\Re\left[ e^{-\im k\log\sqrt{\frac{\lambda}{\lambda_0}}}\J{\im k}{\lambda}\right]}{\sqrt{\hbar\cos\left(k\log\frac{\lambda}{\lambda_0}\right)+\hbar\cosh(k\pi)}}\, ,
\end{eqnarray}
where $A_{sc}(k,\lambda)$ is again a normalised Gaussian centred on some classical values $k_c$ and $\lambda_c$. The parameter $\lambda_0$ is a choice of units and was set to one in the numerics. 

We then calculated the expectation value $\expval{v(t)}_{\Psi_{sc}}$ in this state. In this case, to improve the convergence of numerical integration we chose this state to be very sharply peaked in $k$ so that the integral in $k$ was replaced by its integrand, but included a finite variance $\sigma_\lambda$. More details on the numerical accuracy in $\expval{v(t)}_{\Psi_{sc}}$ can be found in \cite{Gielen2020}. This expectation value was then compared with the classical solution
\begin{equation}
v_c(t)=\sqrt{4\lambda_ct^2 -\frac{\hbar^2k_c^2}{\lambda_c}}\,.
\end{equation}

\begin{figure}[h!]
\centering
\includegraphics[scale=0.4]{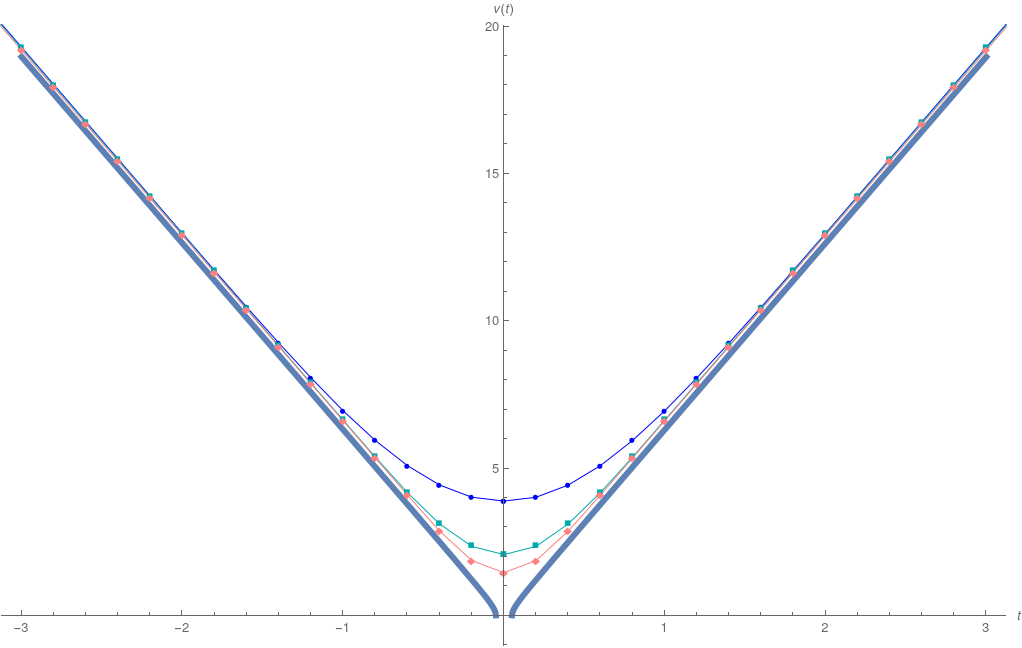}
\caption{Expectation value $\expval{v(t)}_{\Psi_{sc}}$ and classical solution $v_c(t)$ (solid line) for $k=1$ and $\lambda_c=10$, with $\hbar=1$. The blue circles, cyan squares and pink diamonds correspond respectively to the quantum expectation values with $\sigma_\lambda=1,2$ and 3.}
\label{v-t-sigmas}
\end{figure}

Figure \ref{v-t-sigmas} shows the results for $\expval{v(t)}_{\Psi_{sc}}$ for several values of $\sigma_\lambda$. We observe a number of similarities between this theory and the results obtained in section \ref{num-phi}. First, we see that there is a smooth transition between contracting and expanding classical solutions, and classical and quantum solutions agree very well at large values of $\abs{t}$. In section \ref{num-phi} the same was true with respect to  $\abs{\varphi}$. Here we also see the classical and quantum solutions start diverging before reaching $v=0$, which is where boundary conditions were imposed in this theory (rather than at $v=\infty$ as in section \ref{num-phi}). At $t=0$ a local extremum is reached but this time it is not a maximum but a minimum: The universe starts with an infinite volume, contracts until reaching a strictly positive value, and then bounces back to infinite volume. Here of course the difference is that this behaviour is interpreted as resolving the classical singularity, rather than a recollapse which ``avoids'' infinity. The minimum value of the volume is state-dependent, just as we observed for the maximum volume in section \ref{num-phi}. For larger variance $\sigma_\lambda$ one finds smaller $\expval{v(0)}_{\Psi_{sc}}$ and close agreement between classical and quantum solution for a longer time. This mimics exactly what we saw in figure \ref{v-sigmas} when changing $\sigma_k$. 

We see that the three theories behave very differently in the semiclassical regime, as expected due to the different boundary conditions that were imposed in section \ref{comparsec}:
\begin{itemize}
\item \emph{Divergence from the classical theory}. The theories based on $t$ or $\varphi$ as clocks require reflective boundary conditions to be unitary. These theories are then well-behaved for all values of the clocks $t$ and $\varphi$, even the ones that are classically forbidden, which inevitably leads to divergence from the classical theory. On the other hand, if $v$ is used as a clock, the theory is automatically unitary and there is no boundary condition and no strong divergence from the classical theory.

\item \emph{Fast clocks and slow clocks}. One may wonder what distinguishes the boundary value $v=0$ when $t$ is the clock, but $v=\infty$ when $\varphi$ is the clock. Following the terminology used by Gotay and Demaret \cite{Gotay1984,*Gotay1996}, at these points $t$ and $\varphi$ are ``slow'' clocks: for any classical solution $v=0$ is always reached at finite $t$ and $v=\infty$ at finite $\varphi$. We could say that these clocks do not tick ``fast enough'' when reaching these singular points. On the contrary, a ``fast'' clock is one that reaches the singular point only asymptotically at the boundaries of its domain. One clock can be fast at one point and slow at another point: $t$ is slow at $v=0$ but fast at $v=\infty$. In \cite{Gotay1984,*Gotay1996} Gotay and Demaret conjectured that slow (fast) clocks (do not) resolve the singular points, basically because of what we mentioned in the previous paragraph: a slow clock requires boundary conditions, and hence departure from the classically singular solutions. Our results in \cite{Gielen2020} and in this work back up this conjecture: the $t$ clock theory is singularity free and the $\varphi$ clock theory is ``infinity'' free (it avoids $v=\infty$). The $v$ clock is fast at both singular points, as is best seen by noting that the natural clock variable is $\log(v/v_0)$ which goes to $-\infty$ as $v\rightarrow 0$. This quantum theory can then stay close to the classical theory at all times. 

\end{itemize}

A good way to visualise the differences between the quantum theories is to use conformal (Penrose--Carter) diagrams. To construct a conformal diagram explicitly, we would have to apply a suitable conformal transformation to the metric \eqref{metric} to map spacetime into a finite region. However, in our case it is sufficient to consider the asymptotic regimes of large and small volume. Close to the big bang/big crunch singularity, which is spacelike, the dynamics are dominated by the scalar field (assuming, as always, that $\pi_\varphi\neq 0$). Then at large $v$, if we take the perfect fluid to represent dark energy and assume $\lambda>0$, the universe is asymptotically de Sitter. The conformal diagram is obtained by looking at the causal structure in these asymptotic regimes and gluing them together. We only show the conformal diagrams for this dark energy interpretation ($w=-1$) but if we chose a value $w>-1$, the universe would asymptote to Minkowski spacetime. The reason why the conformal structure depends on the choice of equation of state parameter $w$ is that different $w$ corresponds to different choices of lapse in (\ref{Hamiltonian}) if one works e.g.~in the time coordinate $t$.

Our quantum theories are the result of symmetry reduction at the classical level, so the connection between a particular choice of time coordinate and the spacetime metric (as expressed by the lapse) is no longer obvious at the quantum level. Here we assume that a particular time coordinate has the same interpretation in the classical and quantum theories, i.e.~that the form of the lapse is unchanged. This is justified since in any case the conformal diagrams we draw can only represent expectation values in a quantum state, and clearly only make sense in a semiclassical regime. Their main point is to illustrate where the corrections to the classical geometry are small or large.

\begin{figure}
\begin{center}
\begin{tikzpicture}
\node (I) at (-5,0) {I};
\node (II) at (5,0) {II};

\path 
(I)+ (45:4) coordinate (Itr)
+(135:4) coordinate (Itl)
+ (225:4) coordinate  (Ibl)
+(-45:4) coordinate (Ibr);

\draw
 (Itl) -- (Ibl) (Ibr) -- (Itr);

\draw[decorate,decoration=zigzag] (Itr) --node[midway, above] {$v=0$} (Itl);

\draw[line width=2pt] (Ibl)  --node[midway, below] {$v=\infty$} (Ibr);

\path 
(II)+ (45:4) coordinate (IItr)
+(135:4) coordinate (IItl)
+ (225:4) coordinate  (IIbl)
+(-45:4) coordinate (IIbr);

\draw
 (IIbl) -- (IItl)  (IItr) -- (IIbr);

\draw[decorate,decoration=zigzag] (IIbr) --node[midway, below] {$v=0$} (IIbl);

\draw[line width=2pt] (IItl) --node[midway, above] {$v=\infty$} (IItr);
\end{tikzpicture}
\caption{Conformal diagram for classical solutions. The zigzag line represents the singularity and the thicker line represents spacelike infinity ($\mathcal{I}^-$ or $\mathcal{I}^+$). I is a contracting universe and II is an expanding universe. There is no trajectory linking I and II.}
\label{classth-cd}
\end{center}
\end{figure}
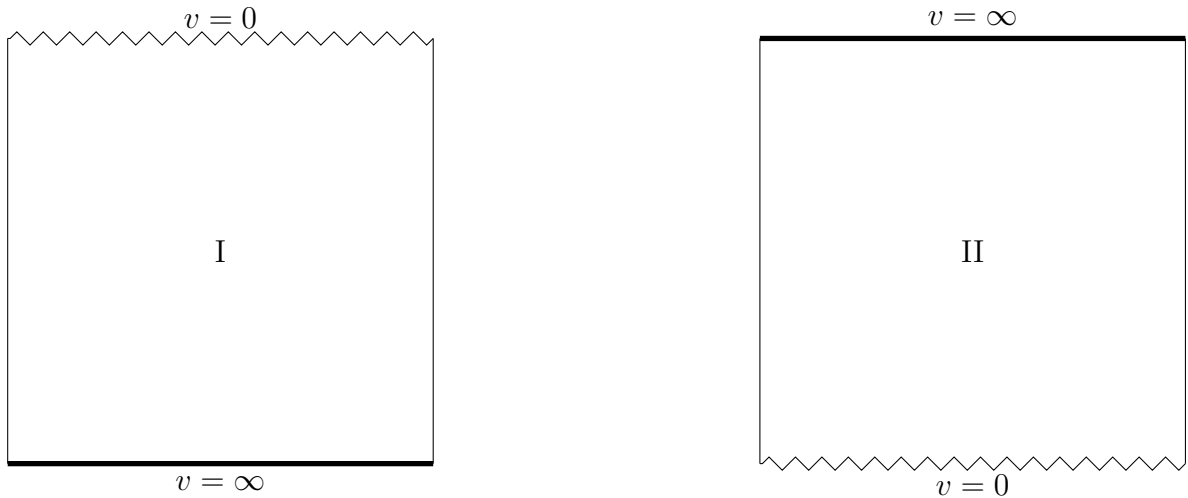

The conformal diagram of the classical theory is represented in figure \ref{classth-cd}. There are two possible solutions, a contracting universe I and an expanding universe II. As usual, light cones are determined by taking lines at $\pm$ 45$^{\mathbf{o}}$, hence any timelike trajectory can be represented by a curve whose tangent vector at any point always stays inside the light cone at that point. In the contracting universe, the singularity is in the future light cone of every observer. The reverse happens in the expanding universe, where the singularity is in the past light cone of every observer. 

We have seen that if $v$ is chosen as the clock, for semiclassical states $\expval{t(v)}_{\Psi_{sc}}$ is very close to the classical $t(v)$ at all times. Therefore, we say that the conformal diagram of this quantum theory is identical to figure~\ref{classth-cd} for semiclassical states $\Psi_{sc}$.

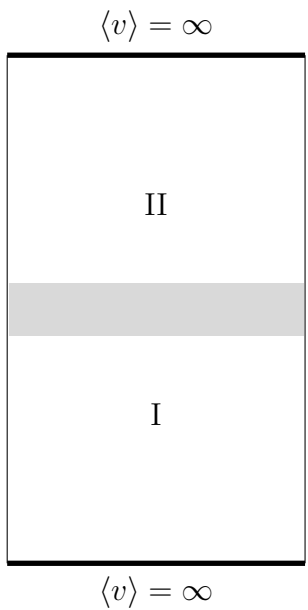
\begin{figure}
\begin{center}
\begin{tikzpicture}[scale=0.7]
\node (I) at (0,-2) {I};
\node (II) at (0,2) {II};

\path 
(I)+ (45:4) coordinate (Itr)
+(135:4) coordinate (Itl)
+ (225:4) coordinate  (Ibl)
+(-45:4) coordinate (Ibr);

\path 
(II)+ (45:4) coordinate (IItr)
+(135:4) coordinate (IItl);

\draw (Ibr)--(Itr)--(IItr) (Ibl)--(Itl)--(IItl);
\draw[line width=2pt] (Ibr)--node[midway, below] {$\expval{v}=\infty$} (Ibl);
\draw[line width=2pt] (IItr)--node[midway, above] {$\expval{v}=\infty$} (IItl);
\fill[gray, fill opacity=0.3] (-2.8,-0.5) rectangle (2.8,0.5);
\end{tikzpicture}
\caption{Conformal diagram of the $t$ clock theory, with contracting region I and expanding region II. The singularity is replaced by the shaded area, where quantum fluctuations are large. In this state-dependent region the volume reaches a minimum expectation value $\expval{v}=V_{min_\Psi}$ and it is impossible to talk about a ``classical trajectory''.}
\label{t-clock-cd}
\end{center}
\end{figure}

The conformal diagram for semiclassical states in the $t$ clock theory is presented in figure \ref{t-clock-cd}; we replace the classical $v(t)$ by the expectation value $\expval{v(t)}_{\Psi_{sc}}$. This expectation value does not go to zero but has a strictly positive minimum value before growing to infinity; accordingly universes I and II from figure \ref{classth-cd} have been glued together, removing the singularity. The classical singularity is replaced by a region in which quantum fluctuations are large, and where classical trajectories are not well-defined. The region near spacelike infinity is identical to the classical theory. 

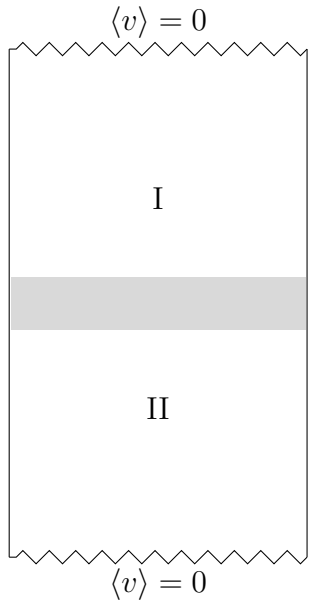
\begin{figure}
\begin{center}
\begin{tikzpicture}[scale=0.7]
\node (I) at (0,-2) {II};
\node (II) at (0,2) {I};

\path 
(I)+ (45:4) coordinate (Itr)
+(135:4) coordinate (Itl)
+ (225:4) coordinate  (Ibl)
+(-45:4) coordinate (Ibr);

\path 
(II)+ (45:4) coordinate (IItr)
+(135:4) coordinate (IItl);

\draw (Ibr)--(Itr)--(IItr) (Ibl)--(Itl)--(IItl);
\draw[decorate,decoration=zigzag] (Ibr)--node[midway, below] {$\expval{v}=0$} (Ibl);
\draw[decorate,decoration=zigzag] (IItr)--node[midway, above] {$\expval{v}=0$} (IItl);
\fill[gray, fill opacity=0.3] (-2.8,-0.5) rectangle (2.8,0.5);
\end{tikzpicture}
\caption{Conformal diagram of the $\varphi$ clock theory. Now the contracting region I lies to the future of the expanding region II. The singularity is still present, but the expectation value of the  volume remains finite. At large $v$ there is a region in which quantum fluctuations dominate and where the volume reaches its maximum expectation value $\expval{v}=V_{max_\Psi}$. Once again, in this region there is no notion of classical trajectory.}
\label{phi-clock-cd}
\end{center}
\end{figure}

The conformal diagram of the $\varphi$ theory (figure \ref{phi-clock-cd}) represents the expectation value $\expval{v(\varphi)}_{\Psi_{sc}}$ instead of the classical $v(\varphi)$. Here the classical singularity remains but infinity is ``resolved'' and so regions I and II are glued in the opposite way: spacelike infinity is replaced by a quantum region but the classical singularity is unchanged. Again there is a region in which quantum fluctuations dominate, represented again by a grey area.

In closing, we mention that the behaviour suggested in figure \ref{phi-clock-cd} has some similarities to Penrose's conformal cyclic cosmology scenario \cite{Penrose}: spacelike infinity is no longer seen as the future endpoint of a $\Lambda$-dominated universe but becomes a transition point into a new universe. Of course the key differences, as already pointed out in \cite{Pawlowski2011}, are that the subsequent ``aeon'' is contracting, not expanding, and that the origin of this transition is highly quantum. The cosmological scenario emerging in our analysis could potentially be turned into a fully cyclic picture if time evolution was controlled by a ``slow'' clock both at singularities and at spacelike infinity; unitarity would then enforce resolution of the singularity, and replacement of spacelike infinity by a quantum recollapse. We leave exploration of this interesting idea to future work. 

\section{Unitarity and self-adjointness criteria}
\label{dirac}

As we discussed at various points in the paper, the quantum recollapse seen in our model both analytically and numerically is due to the boundary condition at $v=\infty$, which itself is a consequence of our requirement of unitarity: we demanded that inner products between states are conserved in $\varphi$ time, $\pdv{}{\varphi}\braket{\Psi}{\Phi}_{\varphi}=0$. For $\lambda>0$, classical solutions terminate at a finite value of $\varphi$ where the volume of the universe diverges, but a quantum solution must continue beyond this point; the quantum state is reflected from $v=\infty$. These results can be intuitively understood by comparing with analogous situations in quantum mechanics, where a strongly attractive potential requires the same type of boundary condition for the Hamiltonian to be self-adjoint.

An obvious question to ask is in which way these results follow from our choice of $\varphi$ as a clock, or more specifically from the particular inner product (\ref{phi-inner-prod}) adapted to $\varphi$. One might  also want to compare our constructions to theories defined via the ``clock-neutral'' approach of Dirac quantisation \cite{Marolf2000,*Marolf1995,Tate1992} in which one first defines a kinematical Hilbert space and defines physical states as solutions to a quantum constraint equation on this Hilbert space; here at least {\em a priori} no choice of clock is necessary. The equivalence of quantum theories defined with respect to different clocks, as well as the equivalence of Dirac quantisation and other relational approaches to quantisation of constrained systems has been established \cite{Hoehn2019,*Hoehn2020} warranting the terminology ``clock-neutral'' for this approach. Because of its more democratic treatment of different clocks this approach is often seen as preferable to the type of quantisation we have used so far, where a particular clock is singled out.

We mentioned earlier that the Dirac quantisation of a model very similar to ours (the only difference being that $\lambda>0$ is a fixed parameter) was studied in \cite{Pawlowski2011}. In this work, the Hamiltonian constraint was required to be self-adjoint on the kinematical Hilbert space, which led to a boundary condition at $v=\infty$ and a recollapse of the universe similar to our results here. The physical inner product used in \cite{Pawlowski2011} is constructed using a group averaging procedure in a representation different from the one used in this paper, but one might expect the resulting inner product to be equivalent to the inner product (\ref{phi-inner-prod}) modulo the issue of positive definiteness which we discussed earlier; in simpler systems such as a relativistic particle group averaging leads to an inner product of Klein--Gordon form (see e.g.~\cite{Hoehn2018b}).

We can try to see explicitly what would happen in a Dirac quantisation of our model if we require the Hamiltonian constraint to be represented as a self-adjoint operator on the kinematical Hilbert space\footnote{Although this is a common assumption, some approaches to Dirac quantisation do not require self-adjointness of a Hamiltonian constraint, e.g.~if the inner product is not constructed through group averaging. (We thank Philipp H\"ohn for pointing this out to us.)}. Let us write the Wheeler--DeWitt equation (\ref{wdw2}) as
\begin{equation}
\hat{\mathcal{C}}_1\Psi(u,\varphi,t)=0\,,\quad\hat{\mathcal{C}}_1:=\hbar^2\frac{\partial^2}{\partial u^2}-\hbar^2\frac{\partial^2}{\partial\varphi^2}-\im\hbar v_0^2 e^{2u} \pdv{}{t}\,,
\label{const1}
\end{equation}
where we now use the variable $u=\log(v/v_0)$ in order for the kinematical variables $u$, $\varphi$ and $t$ to all be valued over the entire real axis. The second derivatives in (\ref{const1}) correspond to the Laplacian on (1+1) dimensional flat space in standard Cartesian coordinates, which motivates defining a kinematical Hilbert space by the inner product
\begin{equation}
\braket{\Psi}{\Phi}_{{\rm kin(1)}}=\int \dd t\;\dd\varphi\; \dd u\;\bar{\Psi}(u,\varphi,t)\Phi(u,\varphi,t)\,,
\label{kine-inner-prod-2}
\end{equation}
i.e. assuming that the metric on the submanifold parametrised by $u$ and $\varphi$ is simply the flat metric $g_{AB}=\eta_{AB}$. It is important to clarify that this is {\em not} the (equally flat) metric $g_{AB}=v_0^2 e^{2u} \eta_{AB}$ on the Rindler wedge that we derived in (\ref{metricformham}), but a conformally rescaled metric. This is because in going from the original Wheeler--DeWitt equation (\ref{wdw}) to its equivalent form (\ref{wdw2}) we multiplied by $v_0^2 e^{2u}$, which corresponds to this conformal rescaling (because we are on a flat geometry, the conformally rescaled Laplacian is the Laplacian of the conformally rescaled metric). The constraint written in the form (\ref{const1}) is indeed the one used in \cite{Pawlowski2011}, up to potential ordering ambiguities.

The requirement on $\hat{\mathcal{C}}_1$ to be self-adjoint with respect to (\ref{kine-inner-prod-2}) does not pose any nontrivial boundary conditions with respect to $\varphi$ or $t$; however in the $u$ variable we can see that it is equivalent to asking that the operator
\begin{equation}
\hat{\mathfrak{O}}_1=-\hbar^2\frac{\partial^2}{\partial u^2}-\lambda e^{2u}
\end{equation}
be self-adjoint on $L^2(\mathbb{R})$ for any value of $\lambda$. This is exactly the condition we already found in our quantisation earlier, see the discussion around (\ref{fr-ham}); hence the boundary condition at $u=\infty$ will be the one we discussed earlier in this paper.

What happens if we write the Wheeler--DeWitt equation in the original form (\ref{wdw})? In this case, for
\begin{equation}
\hat{\mathcal{C}}_2\Psi(u,\varphi,t)=0\,,\quad\hat{\mathcal{C}}_2:=\hbar^2\frac{\partial^2}{\partial v^2}+\frac{\hbar^2}{v}\pdv{}{v}-\frac{\hbar^2}{v^2}\frac{\partial^2}{\partial\varphi^2}-\im\hbar \pdv{}{t}\,,
\label{const2}
\end{equation}
we would ask that $\hat{\mathcal{C}}_2$ is self-adjoint with respect to an appropriate kinematical inner product, which we now take as
\begin{equation}
\braket{\Psi}{\Phi}_{{\rm kin(2)}}=\int_{-\infty}^{\infty} \dd t\;\dd\varphi\int_0^\infty \dd v\;v\;\bar{\Psi}(u,\varphi,t)\Phi(u,\varphi,t)\,.
\label{kine-inner-prod-3}
\end{equation}
In this inner product the measure has an extra term $\sqrt{-g}=v$ since we now use the metric (\ref{metricformham}) whose Laplacian appears in (\ref{const2}). Wavefunctions that are square-integrable  in this inner product are of the form $\Psi=v^{-1/2}\Xi$ where $\Xi$ is square-integrable in a standard $L^2$ inner product (with trivial measure) on $\mathbb{R}^2\times\mathbb{R}_+$. One then sees that self-adjointness of $\hat{\mathcal{C}}_2$ with respect to the inner product (\ref{kine-inner-prod-3}) is equivalent to demanding that the operator
\begin{equation}
\hat{\mathfrak{O}}_2=\hbar^2\left(-\frac{\partial^2}{\partial v^2}-\frac{k^2+\frac{1}{4}}{v^2}\right)
\end{equation}
is self-adjoint on $L^2(\mathbb{R}_+, \dd v)$. But this is exactly the condition we previously found for unitarity with respect to the clock $t$, see section \ref{comparsec}! Indeed, in this case it is even clearer that the Schr\"odinger-type quantisation studied in our previous paper \cite{Gielen2020} is equivalent to a Dirac quantisation for the constraint (\ref{const2}), given that the constraint is linear in the momentum $\lambda$ conjugate to $t$ and so the usual group averaging procedure would only remove the integral over $t$ from the measure (\ref{kine-inner-prod-3}), leading to the Schr\"odinger-type inner product studied in \cite{Gielen2020} and previously in \cite{Gryb2019,Gryb2019b,*Gryb2019c}.

We then see that the different notions of unitarity with respect to different clocks do not arise from using different inner products adapted to different clocks, but actually from writing the Wheeler--DeWitt equation either in the original form (\ref{const2}) or in the form (\ref{const1}) where one has multiplied the equation by a nontrivial phase-space function. From the perspective of the classical theory, these different Wheeler--DeWitt equations correspond to different choices of lapse: in the discussion of section \ref{classth} we wrote the Hamiltonian constraint as
\begin{equation}
\mathcal{H} = \tilde{N} \left[-\pi_v^2+\frac{\pi_\varphi^2}{v^2}+\lambda\right]
\end{equation}
so that (\ref{const2}) corresponds to choosing $\tilde{N}=1$ whereas (\ref{const1}) corresponds to $\tilde{N}=v^2$. Of course, classically any choice of lapse function is equally valid, with different choices corresponding simply to different time coordinates. Again, we see that this classical symmetry of time reparametrisation is broken in the quantum theory in a subtle way: changing the lapse leads to a different kinematical inner product and then to different criteria for self-adjointness of the Hamiltonian constraint, different boundary conditions and ultimately different physical behaviour, as we saw explicitly earlier. 

Our results are then also not a contradiction to the findings of \cite{Hoehn2019,*Hoehn2020} regarding the equivalence of different clock choices: in these papers the Hamiltonian constraint and the kinematical Hilbert space are taken as given, whereas we are comparing settings in which (from the perspective of Dirac quantisation) one has changed both the measure in the kinematical inner product and the constraint from an initial constraint $\hat{\mathcal{C}}$ to $ \hat{\mathcal{N}}\hat{\mathcal{C}}$. In the same inner product, these two constraints could both be self-adjoint at the same time if one chose to quantise $\mathcal{N C}$ as $\hat{(\sqrt{\mathcal{N}})}\hat{C}\hat{(\sqrt{\mathcal{N}})}$, which is not what we have done here (and, given that $\sqrt{\mathcal{N}}=v_0 e^u$ does not obviously translate to a self-adjoint operator, might not work anyway). Our point is that $\hat{\mathcal{C}}\Psi=0$ and $\hat{\mathcal{N}}\hat{\mathcal{C}}\Psi=0$ have the same solutions, but live in two different physical Hilbert spaces and define very different theories once unitarity is imposed. It is not clear to us how such a conclusion could be avoided if one maintains the classical reparametrisation symmetry which allows multiplying a classical constraint by any nontrivial phase-space function.

\section{Conclusion}
\label{ccl}
In this paper we have expanded our analysis in \cite{Gielen2020} of a spatially flat FLRW universe with an arbitrary perfect fluid and a free massless scalar field by studying a quantisation based on the relational clock $\varphi$ and comparing the resulting quantum theory to the previously analysed theories based on the clocks $t$ and $v$. We had previously found a lack of general covariance when comparing different quantum theories, and in this paper we provide even more evidence of this fact. Perhaps the main insight of this new work is how imposing unitarity as a fundamental principle leads to the appearance of boundary conditions, and the divergence from the classical theory. We have reproduced and extended some previous results for the quantum dynamics of the same model with the same clock, e.g.~\cite{Bojowald2010b, Pawlowski2011}. We have also verified Gotay and Demaret's conjecture \cite{Gotay1984,*Gotay1996} for our model and expanded its scope to predict not only singularity resolution but also a quantum recollapse of the universe. If we assume its more general validity, this conjecture could be particularly useful when choosing a relational clock for other quantum cosmological models: it would suggest that a clock that is slow (fast) at the singularity or at infinity will (will not) resolve the singularity or will (will not) lead to a quantum recollapse. All that would be required to achieve singularity resolution, in {\em any} model of quantum cosmology, would be to add a degree of freedom that remains finite at the singularity, as is the case for the clock $t$ in our model. Of course, this remains a conjecture which needs to be substantiated with further evidence.

In order to construct the quantum theories we studied, we had to make some assumptions. First of all, when deriving the Wheeler--DeWitt equation \eqref{wdw}, we imposed covariance under a change of coordinate in $(v,\varphi)$ in order to fix the operator ordering. With a different ordering the details of our results might change. A second choice was that of an inner product, which we fix in the usual way by taking the quadratic part of the constraint to define a metric on minisuperspace. This procedure indirectly introduces a dependence on the lapse function $N$, since multiplying the constraint by a nontrivial function would change this minisuperspace metric. In this sense, the choice of lapse does not affect the solutions of the Wheeler--DeWitt equations but changes the inner product. All of these ambiguities are well-known in the field of quantum cosmology. Still, the main assumption of this paper is that we impose unitarity as a fundamental principle. Imposing unitarity is then equivalent to imposing self-adjointness of an operator which is not originally self-adjoint, and one has to choose one of the possible self-adjoint extensions. In our case this means choosing a free function of one variable. In this paper we have not studied in detail how the choice of self-adjoint extension affects the dynamics of the theory. In \cite{Pawlowski2011} Paw{\l}owski and Ashtekar found that the specific self-adjoint extension does not seem to play a significant role. 

One might of course assume that unitarity (or self-adjointness of a Hamiltonian) is not a fundamental property of quantum gravity. For instance, one of the possible approaches to the problem of time is to see time and unitarity as emergent concepts that are only well-defined semiclassically, see e.g.~\cite{Isham1992,Vilenkin1989,*Kiefer1991,*DiGioia2021}, and to not demand any fundamentally conserved inner product. The absence of unitarity, in the context of our models, would presumably imply that the norm of semiclassical states goes to zero as they approach the singularity, which would be hard to interpret. To us it would seem difficult to motivate dropping such a fundamental principle of quantum mechanics in quantum gravity while keeping other principles such as the standard canonical quantisation of the classical constraint leading to the Wheeler--DeWitt equation.

We explained why our results are not in contradiction with the results of H\"ohn and collaborators \cite{Hoehn2019,*Hoehn2020} regarding covariant Dirac quantisation: while the dynamics for different clocks are based on the same Wheeler--DeWitt equation, the self-adjointness requirement usually imposed to construct the inner product applies to different constraint operators, which correspond to different choices of lapse. Moreover, the kinematical inner product would also be different for different clocks. The same issue we already mentioned above, namely the dependence on the choice of lapse, hence also arises in this more covariant approach to canonical quantisation. 

In \cite{Pawlowski2011} our model with fixed $\lambda>0$ was studied first in the Wheeler--DeWitt approach we are using here but then also in loop quantum cosmology. In the latter case, one finds that the singularity was resolved using the $\varphi$ clock but the same self-adjoint extension problem as in the Wheeler--DeWitt approach results from boundary conditions at infinity. It would be interesting to study in general which aspects of the dynamics in quantum cosmology depend on the choice of self-adjoint extension used and if other relational clocks can be used in loop quantum cosmology in particular. 

Another important direction for future work would be to study whether one could overcome the issue of clock dependence of the quantum theory by using a path integral quantisation. In that formalism there are techniques that ensure gauge invariance \cite{GaugeSystems}, and  the path integral can at least be formally defined in a way that is independent of the choice of lapse \cite{Halliwell1988}. 

Perhaps more philosophically, we might ask ourselves if it is true that every monotonic classical variable is really a good clock. For example, in this model the $\varphi$ clock, although \emph{mathematically} a good clock, might not be \emph{physically} as well motivated: according to this clock the singularity is infinitely far away, but infinity is reached in a finite time. No classical timelike observer will ever \emph{experience} the passing of time in that way. In the late universe, an observer carrying a clock device measuring the value of the scalar field would see how its clock freezes. Should we limit ourselves to clocks that potentially measure proper time for an observer? In a way time is not only what clock measures but what observers experience. 

One of the main motivations for studying quantum rather than classical
cosmology is the hope that quantum effects will cure the
singularities of classical general relativity, in particular the big
bang. Observing nonclassical behaviour is hence essential if quantum
cosmology is to provide us with new insights about gravity in extreme
regimes. In our context, such departures from classical dynamics arise
from the reflecting behaviour at classical singularities implied by
demanding unitarity. This reflection can represent a type of singularity
resolution, but only if we use a clock that is slow at the singularity.
We saw that a clock can also be slow at infinity, which will again
trigger nonclassical behaviour but now in a regime where curvature is
very low and one would not expect quantum gravity to be relevant. We
observed this counterintuitive behaviour in this paper for the scalar
field clock; taken at face value it would imply a dramatic infrared
modification of classical general relativity caused by large quantum
fluctuations at very large volume. To us, the fact that all of these
``predictions'' depend on which clock variable is used raises
more questions than it answers. It seems that the problem of time must
be addressed, and a more fully covariant quantisation be understood,
before we can trust predictions of departure from classical physics in
quantum cosmology.

\section*{Acknowledgments}
We would like to thank Philipp H\"ohn for helpful comments on our work, Jorma Louko for comments on the manuscript and on calculations involving Bessel functions in Appendix A, and a referee for important technical comments. The work of SG was funded by the Royal Society through a University Research Fellowship (UF160622) and a Research Grant for Research Fellows (RGF\textbackslash R1\textbackslash 180030).

\appendix

\section{Important integrals}
\label{calculations}

In this appendix we collect the results of various integrals involving two Bessel functions. These integrals are used in section \ref{qtumth} when computing the inner product (\ref{phi-inner-prod}) of two solutions to the Wheeler--DeWitt equation and in section \ref{numres} when computing expectation values in the same inner product.

The integrals of interest in section \ref{qtumth} are of the form
\begin{equation}
\mathcal{J}_{\mu,\nu}:=\int_0^\infty \frac{{\rm d}v}{v}\,J_{\mu}( C v)J_{\nu}(C v)\,,\quad \mathcal{K}_{\mu,\nu}:=\int_0^\infty \frac{{\rm d}v}{v}\,K_{\mu}( C v)K_{\nu}(C v)
\end{equation}
where the orders $\mu$ and $\nu$ are either real or imaginary. These integrals are independent of the positive parameter $C$ on the right-hand side.

The integrals of interest in section \ref{numres} are of the form
\begin{equation}
\mathcal{O}_{\mu,\nu,C}:=\int_0^\infty {\rm d}v\,J_{\mu}(C v)J_{\nu}(C v)
\end{equation}
where $\mu$ and $\nu$ are either imaginary or of the form $\pm 1 + \im\alpha$ where $\alpha\in\mathbb{R}$ and $C>0$. We present these integrals in the order in which they appear in the main text.

\subsection{The integrals $\mathcal{J}_{\im\alpha,\im\beta}$ and $\mathcal{K}_{\im\alpha,\im\beta}$}

These integrals appear in computing orthogonality relations between different states.
The parameters $\alpha$ and $\beta$ are real. The integral $\mathcal{J}_{{\rm i}\alpha,{\rm i}\beta}$ does not converge but it can be defined in a
distributional sense as a limit of the integral
\newpage
\begin{eqnarray}
\fl
&& \lim_{\nu\rightarrow 1}\int_0^\infty \frac{{\rm d}x}{x^\nu}J_{{\rm i}\alpha}(x)J_{{\rm i}\beta}(x)\nonumber
\\
\fl
&=&\lim_{\nu\rightarrow 1}\frac{2^{-\nu}\Gamma\left(\frac{1-\nu}{2}+\frac{{\rm i}}{2}(\alpha+\beta)\right)\Gamma(\nu)}{\Gamma\left(\frac{1+\nu}{2}+\frac{{\rm i}}{2}(\alpha-\beta)\right)\Gamma\left(\frac{1+\nu}{2}+\frac{{\rm i}}{2}(\beta-\alpha)\right)\Gamma\left(\frac{1+\nu}{2}+\frac{{\rm i}}{2}(\alpha+\beta)\right)}
\end{eqnarray}
which is initally only defined for $\nu<1$; notice the possible singularity in the first Gamma function in the numerator as $\nu\rightarrow 1$. We have also introduced the integration variable $x=Cv$ to simplify the notation in this integral. To proceed, we can now rewrite
\begin{equation}
\Gamma\left(\frac{1-\nu}{2}+\frac{{\rm i}}{2}(\alpha+\beta)\right)=\frac{\Gamma\left(\frac{3-\nu}{2}+\frac{{\rm i}}{2}(\alpha+\beta)\right)}{\frac{1-\nu}{2}+\frac{{\rm i}}{2}(\alpha+\beta)}
\label{gammaplusone}
\end{equation}
so that we obtain
\begin{equation}
\fl
\lim_{\nu\rightarrow 1}\int_0^\infty \frac{{\rm d}x}{x^\nu}J_{{\rm i}\alpha}(x)J_{{\rm i}\beta}(x) =  \frac{2\sinh\left((\alpha-\beta)\frac{\pi}{2}\right)}{\pi(\alpha-\beta)}\times\lim_{\nu\rightarrow 1}\frac{1}{(1-\nu+{\rm i}(\alpha+\beta))}\,.
\end{equation}
The last limit must now be taken in a distributional sense using the identity
\begin{equation}
\lim_{\epsilon\rightarrow 0^+}\frac{1}{y+{\rm i}\epsilon}={\rm PV}\frac{1}{y}-{\rm i}\pi\delta(y)
\end{equation}
where ${\rm PV}$ denotes the Cauchy principal value, i.e.~the distribution defined by
\begin{equation}
\int_{-\infty}^\infty {\rm d}y\left[{\rm PV}\frac{1}{y}\right]f(y) = \int_0^\infty \frac{{\rm d}y}{y}(f(y)-f(-y))
\end{equation}
for any test function $f(y)$, which depends only on the odd part of $f$. In summary we then find
\begin{equation}
\fl
\int_0^\infty \frac{{\rm d}x}{x}J_{{\rm i}\alpha}(x)J_{{\rm i}\beta}(x) =  2\sinh\left((\alpha-\beta)\frac{\pi}{2}\right)\left(\frac{\delta(\alpha+\beta)}{\alpha-\beta}-{\rm PV}\frac{{\rm i}}{\pi(\alpha^2-\beta^2)}\right)\,.
\label{2-Bessel-int-1}
\end{equation}
In the case of modified Bessel functions we can proceed in the same fashion;  we find
\begin{equation}
\fl
\lim_{\nu\rightarrow 1}\int_0^\infty \frac{{\rm d}x}{x^\nu}K_{{\rm i}\alpha}(x)K_{{\rm i}\beta}(x)=\lim_{\nu\rightarrow 1}\frac{\left|\Gamma\left(\frac{1-\nu}{2}+\frac{{\rm i}}{2}(\alpha-\beta)\right)\right|^2\left|\Gamma\left(\frac{1-\nu}{2}+\frac{{\rm i}}{2}(\alpha+\beta)\right)\right|^2}{2^{2+\nu}\Gamma(1-\nu)}
\end{equation}
which has a more complicated singularity structure, with possible singularities in all Gamma functions. By substitutions similar to (\ref{gammaplusone}) we obtain
\begin{eqnarray}
\lim_{\nu\rightarrow 1}\int_0^\infty \frac{{\rm d}x}{x^\nu}K_{{\rm i}\alpha}(x)K_{{\rm i}\beta}(x) &=& \frac{\pi^2(\alpha^2-\beta^2)}{\cosh(\alpha\pi)-\cosh(\beta\pi)}\times
\\&&\lim_{\nu\rightarrow 1}\frac{(1-\nu)}{|1-\nu+{\rm i}(\alpha-\beta)|^2|1-\nu+{\rm i}(\alpha+\beta)|^2}\,.\nonumber
\end{eqnarray}
If we now exclude the case $\alpha=\beta=0$, then at least one of the two factors in the denominator remains regular as $\nu\rightarrow 1$ and can be taken outside of the limit. For the second factor we have to take the distributional limit
\begin{equation}
\lim_{\epsilon\rightarrow 0^+}\frac{\epsilon}{\epsilon^2+y^2}=\pi\delta(y)
\end{equation}
as can be seen from
\begin{equation}
\lim_{\epsilon\rightarrow 0^+}\int_{-\infty}^\infty {\rm d}y\;f(y)\;\frac{\epsilon}{\epsilon^2+y^2} = \lim_{\epsilon\rightarrow 0^+}\int_{-\infty}^\infty {\rm d}\upsilon\;f(\epsilon\upsilon)\;\frac{1}{1+\upsilon^2}=\pi f(0)
\end{equation}
where $f$ is again a test function. Altogether we have
\begin{eqnarray}
\int_0^\infty \frac{{\rm d}x}{x}K_{{\rm i}\alpha}(x)K_{{\rm i}\beta}(x) &=& \frac{\pi^3(\alpha^2-\beta^2)}{\cosh(\alpha\pi)-\cosh(\beta\pi)}\left(\frac{\delta(\alpha-\beta)}{(\alpha+\beta)^2}+\frac{\delta(\alpha+\beta)}{(\alpha-\beta)^2}\right)\nonumber
\\&=& \frac{\pi^2}{2\alpha\sinh(\alpha\pi)}\left(\delta(\alpha-\beta)+\delta(\alpha+\beta)\right)\,.
\label{kintegral}
\end{eqnarray}
Notice that the modified Bessel functions of the second kind are always real even for imaginary order, hence there is no imaginary contribution leading to a principal value. Such imaginary contributions come from the large $x$ limit of the integral, whereas the right-hand side of (\ref{kintegral}) only comes from the lower limit $x=0$.

\subsection{The integrals $\mathcal{J}_{a,b}$ and $\mathcal{K}_{a,b}$}
Again $a$ and $b$ are real numbers. Here, in order to identify the cases where the integral can be defined, we first evaluate the indefinite integral
\begin{eqnarray}
&&\int_{x_1}^{x_2} \frac{{\rm d}x}{x}\,J_{a}(x)J_{b}(x) \\
& =& \left[\frac{x \left(J_{a-1}(x)J_{b}(x) - J_{a}(x)J_{b-1}(x)\right)}{a^2-b^2}-\frac{\,J_{a}(x)J_{b}(x)}{a+b} \right]_{x=x_1}^{x=x_2}\nonumber
\end{eqnarray}
where we again defined $x=C v$  for simplicity. 
After now substituting the large argument and small argument asymptotic expressions of the Bessel functions we find
\begin{equation}
\int_{0}^{\infty} \frac{\dd x}{x} J_{a}(x) J_{b}(x)= 2\frac{\sin((a-b)\frac{\pi}{2})}{(a^2-b^2)\pi}-\lim_{x\rightarrow  0} \frac{(x/2)^{a+b}}{(a+b)\Gamma(1+a)\Gamma(1+b)}.
\end{equation}
We now see that the integral is finite when $a+b>0$; otherwise the second term makes the integral divergent and undefinable even in a distributional sense. For $a+b>0$,
\begin{equation}
\int_{0}^{\infty} \frac{\dd x}{x} J_{a}(x) J_{b}(x)= 2\frac{\sin((a-b)\frac{\pi}{2})}{(a^2-b^2)\pi}
\label{2-Bessel-int-2}
\end{equation}
which is the standard formula given, for example, as Equation 6.574.2 in \cite{Integrals}. 
\\For the integral $\mathcal{K}_{a,b}$, there is no contribution from large $v$ where the integral falls off but from $v=0$ we find
\begin{eqnarray}
\fl
\int_{0}^{\infty} \frac{\dd x}{x} K_{a}(x) K_{b}(x) &=& -\frac{\pi^2}{4\sin( a\pi)\sin(b\pi)}\times\nonumber
\\&&\lim_{x\rightarrow  0}\left[\frac{(x/2)^{a+b}}{(a+b)\Gamma(1+a)\Gamma(1+b)}-\frac{(x/2)^{-(a+b)}}{(a+b)\Gamma(1-a)\Gamma(1-b)}\right.\nonumber
\\&&+\left.\frac{(x/2)^{b-a}}{(a-b)\Gamma(1-a)\Gamma(1+b)}-\frac{(x/2)^{a-b}}{(a-b)\Gamma(1+a)\Gamma(1-b)}\right]
\label{2-KBessels-int-2}
\end{eqnarray}
so that this integral always diverges for any $a$ or $b$ (this is true also for the case in which $a$ or $b$ are integer, which we do not discuss in detail here).

\subsection{The integrals $\mathcal{J}_{a,\im\beta}$ and $\mathcal{K}_{a,\im\beta}$}
This is the third possible case in which one order is real and the other one is imaginary. This integral appears when computing cross-terms in the inner product in section \ref{qtumth}. In this case the expression resulting from computing first the indefinite integral is
\begin{equation}
\fl
\int_0^{\infty} \frac{\dd x}{x}J_{a}(x)J_{\im\beta}(x)=2\frac{\sin((a-\im\beta)\frac{\pi}{2})}{(a^2+\beta^2)\pi}-\lim_{x\rightarrow 0}\frac{(x/2)^{a+\im \beta}}{(a+\im \beta)\Gamma(1+a)\Gamma(1+\im \beta)}.
\end{equation}
As $x\rightarrow 0$, the exponential function in the second term has a growing (if $a<0$) or decreasing (if $a>0$) absolute value. If $a>0$, the limit when $x\rightarrow 0$ is 0, making the integral converge to the value
\begin{equation}
\int_0^{\infty} \frac{\dd x}{x}J_{a}(x)J_{\im\beta}(x)=2\frac{\sin((a-\im\beta)\frac{\pi}{2})}{(a^2+\beta^2)\pi}.
\label{2-Bessel-int-3}
\end{equation}
For $a<0$ the integral is divergent. 

The integral $\mathcal{K}_{a,\im\beta}$ is again found to diverge for all real values of $a$.

\subsection{The integral $\mathcal{O}_{\im\alpha,\im\beta,C}$}

Here again $\alpha$ and $\beta$ are real numbers. This integral depends non-trivially on the value of $C$. We use the same method of first evaluating the integral for arbitrary limit values, where it yields
\begin{eqnarray}
\fl
\int_{v_1}^{v_2} {\rm d}v\,J_{\im\alpha}(C v)J_{\im\beta}(C v)&=&-\left[\frac{\im\,v \exp\left(\im(\alpha+\beta)\log\frac{C v}{2}\right)}{(\alpha+\beta-\im)\Gamma(1+\im\alpha)\Gamma(1+\im\beta)}\right.\times
\\&& \fl \left.{}_3 F_4 \left(\frac{1}{2}+\iota,\frac{1}{2}+\iota,1+\iota;1+\im\alpha,\frac{3}{2}+\iota,1+\im\beta,1+2\iota;-C^2 v^2\right)\right]_{v=v_1}^{v=v_2}\nonumber
\end{eqnarray}
which can only be given in terms of generalised hypergeometric functions and where we have defined $\iota:=\im\frac{\alpha+\beta}{2}$. This complicated expression simplifies as $v_1\rightarrow 0$ and $v_2\rightarrow \infty$. First, note that the generalised hypergeometric function defines a power series in $(-C^2 v^2)$ and goes to 1 at $v=0$; because of the additional factor $v$ the contribution from the lower limit vanishes as $v_1\rightarrow 0$.

Using the large $v$ asymptotics of the generalised hypergeometric function we then have, formally,
\begin{eqnarray}
\fl\int_{0}^{\infty} {\rm d}v\,J_{\im\alpha}(Cv)J_{\im\beta}(Cv) &=& \hspace{-2mm} -\frac{\cosh\left((\alpha-\beta)\frac{\pi}{2}\right)}{2\pi C}\hspace{-1mm}\times\lim_{v\rightarrow\infty}\left\{\log\left(\frac{4}{C^2 v^2}\right)+\psi\left(\frac{1+\im(\beta-\alpha)}{2}\right)\right.\nonumber
\\  && \left.+\psi\left(\frac{1+\im(\alpha-\beta)}{2}\right)+2\psi\left(\frac{1+\im(\alpha+\beta)}{2}\right)+2\gamma\right\}
\label{digamma_integral}
\end{eqnarray}
where $\psi$ is the digamma function and $\gamma$ is the Euler--Mascheroni constant.  (\ref{digamma_integral}) diverges logarithmically at large $v$; when using it for numerical evaluation of expectation values, we take the upper limit to some large cutoff value $\Xi_v$ and verify that the final result after integrating over the other variables is not too sensitive to the choice of $\Xi_v$.

\subsection{The integral $\mathcal{O}_{\pm 1+\im\alpha,\im\beta,C}$}
Again we start by evaluating the indefinite integral which yields
\begin{eqnarray}
\fl
\int_{v_1}^{v_2} {\rm d}v\,J_{\pm 1+ \im\alpha}(C v)J_{\im\beta}(C v)&=&\left[\frac{v \exp\left((\pm 1+\im(\alpha+\beta))\log\frac{C v}{2}\right)}{((1\pm 1)+\im(\alpha+\beta))\Gamma(1\pm 1 +\im\alpha)\Gamma(1+\im\beta)}\right.\times
\\&& \fl \left.{}_3 F_4 \left(\frac{1}{2}+\iota',\frac{1}{2}+\iota',1+\iota';1\pm 1+\im\alpha,\frac{3}{2}+\iota',1+\im\beta,1+2\iota';-C^2 v^2\right)\right]_{v=v_1}^{v=v_2}\nonumber
\end{eqnarray}
where now $\iota':=\pm\frac{1}{2}+\im\frac{\alpha+\beta}{2}$. The generalised hypergeometric function goes to 1 at $v=0$. For the case of the minus sign we now have a nontrivial contribution from $v=0$, as we can write (see the discussion around (\ref{phi-infty}))
\begin{eqnarray}
&&\lim_{v\rightarrow 0}\frac{2\im\exp\left(\im(\alpha+\beta)\log\frac{C v}{2}\right)}{C(\alpha+\beta)\Gamma(\im\alpha)\Gamma(1+\im\beta)}\nonumber
\\& = & \lim_{u\rightarrow \infty}\frac{2\im\exp\left(-\im(\alpha+\beta)u\right)}{C(\alpha+\beta)\Gamma(\im\alpha)\Gamma(1+\im\beta)} = \frac{2\,\im}{C} \sinh(\alpha\pi)\delta(\alpha+\beta)\,.
\end{eqnarray}
There is no contribution from the lower limit for the case of the plus sign.
\\Including the contribution from large $v$ we find, again formally,
\begin{eqnarray}
\fl\int_{0}^{\infty} {\rm d}v\,J_{1+\im\alpha}(Cv)J_{\im\beta}(Cv) &=& \hspace{-2mm} -\frac{\cosh\left((\alpha-\beta-\im)\frac{\pi}{2}\right)}{2\pi C}\hspace{-1mm}\times\lim_{v\rightarrow\infty}\left\{\log\left(\frac{4}{C^2 v^2}\right)+\psi\left(\frac{\im(\beta-\alpha)}{2}\right)\right.\nonumber
\\  && \left.+\psi\left(1+\frac{\im(\alpha-\beta)}{2}\right)+2\psi\left(1+\frac{\im(\alpha+\beta)}{2}\right)+2\gamma\right\}
\label{H-int-1}
\end{eqnarray}
and
\begin{eqnarray}
\fl\int_{0}^{\infty} {\rm d}v\,J_{-1+\im\alpha}(Cv)J_{\im\beta}(Cv) &=& \hspace{-2mm} -\frac{\cosh\left((\alpha-\beta+\im)\frac{\pi}{2}\right)}{2\pi C}\hspace{-1mm}\times\lim_{v\rightarrow\infty}\left\{\log\left(\frac{4}{C^2 v^2}\right)+\psi\left(1+\frac{\im(\beta-\alpha)}{2}\right)\right.\nonumber
\\  && \left.+\psi\left(\frac{\im(\alpha-\beta)}{2}\right)+2\psi\left(\frac{\im(\alpha+\beta)}{2}\right)+2\gamma\right\}\nonumber
\\ && + \frac{2\,\im}{C} \sinh(\alpha\pi)\delta(\alpha+\beta)\,.
\label{H-int-2}
\end{eqnarray}
These integrals again diverge logarithmically at large $v$ so we need to cut them off at a fixed cutoff value $\Xi_v$. However, for the calculation of interest in the main text we find that the sum over various integrals $\mathcal{O}_{\pm 1+\im\alpha,\im\beta,C}$ leads to an expression in which all the logarithm terms cancel, and which is hence well-defined in the limit $v\rightarrow\infty$.

\section*{References}

\bibliographystyle{iopart-num}
\bibliography{bib2} 
\end{document}